\documentstyle[12pt]{article}
\begin{document}

\begin{center}

{\large\bf Meson masses within the model
of induced nonlocal quark currents.}\\

{Ja.V.~Burdanov$^1$, G.V.~Efimov$^2$,
S.N.~Nedelko$^2$ and S.A.~Solunin$^1$}\\
{\footnotesize\it $^1$Ivanovo State University, Department of Theoretical
Physics, 153377  Ivanovo, Russia}\\
{\footnotesize\it $^2$Bogoliubov Laboratory of Theoretical Physics,
Joint Institute  for Nuclear Research, 141980  Dubna,
Moscow Region, Russia}
\end{center}

\begin{abstract}

The model of induced quark currents formulated in our recent paper
(Phys. Rev. D51, 174) is developed.
The model being a kind of nonlocal extension of the bosonization procedure
is based on the hypothesis that the QCD vacuum is realized by
the (anti-)self-dual homogeneous gluon field.
This vacuum field provides the analytical quark confinement.
It is shown that a particular form of nonlocality of the quark and gluon
propagators determined by the vacuum field, an interaction of quark spin
with the vacuum gluon field and a localization of meson field at the
center of masses of two quarks can explain the
distinctive features of meson spectrum:
Regge trajectories of radial and orbital excitations,
mass splitting between pseudoscalar and vector mesons,
the asymptotic mass formulas
in the heavy quark limit: $M_{Q\bar Q}\to 2m_Q$
for quarkonia and $M_{Q\bar q}\to m_Q$ for heavy-light mesons.
With a minimal set of parameters (quark masses, vacuum field strength
and the quark-gluon coupling constant) the model describes
to within ten percent inaccuracy
the masses and weak decay constants of mesons from all qualitatively
different regions of the spectrum.\\
PACS number(s): 12.39.-x, 11.10.Lm, 12.38.Aw, 14.40.Gx

\end{abstract}

\section{Introduction}

Achievements of the Nambu-Jona-Lasinio (NJL) model in
description of meson masses, decay constants and so on are
well-known ~\cite{klevansky,hatsuda,eguchi}. This success can
be explained by the bosonization procedure which makes possible to extract
collective modes and dynamical breaking of
SU$_{\rm L}(3)\times$SU$_{\rm R}(3)$ and U$_{\rm A}$(1) symmetries.
At the same time, an incorporation of quark confinement into consideration,
description of heavy quarkonia and heavy-light mesons, radial and
angular excitations of mesons, as well as different form-factors
require an essential modification of the NJL model (e.g.,
see \cite{rob}). Another general disadvantage is the
nonrenormalizability of the local four-fermion interaction.

In recent paper~\cite{efin} we have suggested a model that
in some sense can be considered as an extension of
the standard NJL model. There are two crucial modifications.
First, our model is based on
the hypothesis that the QCD vacuum
to be realized by the (anti-)self-dual homogeneous background gluon field.
Second, the effective quark-quark coupling
is described by the nonlocal four-quark interaction
induced by the one-gluon exchange in presence of the (anti-)self-dual
homogeneous gluon vacuum field.
This vacuum field ensures the analytical quark confinement
and breaks the chiral symmetry.
The model of induced quark currents gives
a basis for investigating of all the above-mentioned problems
>from a general point of view. The main features of the model
are as follows~\cite{efin}.

-- There is the quark confinement. The quark propagator being
an entire analytical function in the complex momentum plane~\cite{leutw1}
has the standard local ultraviolet behavior in the Euclidean region,
and is modified essentially in the physical, i.e., Minkowski region.
In contrast to purely chromomagnetic and chromoelectric configurations,
the (anti-)self-dual homogeneous field is a stable
configuration~\cite{leutw2}.

-- The one-gluon exchange is decomposed into an infinite
sum of current-current interaction terms, in which the quark currents
are nonlocal, colorless and carry a complete set of quantum numbers
including the orbital and radial ones. This effective quark-quark
interaction generates a superrenormalizable perturbation expansion.

-- The bosonization of the nonlocal four-quark interaction
leads to ultraviolet finite effective meson theory. Mesons are
treated as extended nonlocal objects.

-- The model contains the minimal number of parameters: the quark masses,
quark-gluon coupling constant and the tension of the background gluon field.

It was shown also, that the spectrum of the radial
and orbital excitations is equidistant for sufficiently large angular
momentum $\ell$ or radial quantum number $n$. In the heavy quark limit
the mass of quarkonium tends to be equal to sum of the masses of
constituent quarks.

In paper~\cite{efin} the main attention was paid to mathematical details
of obtaining nonlocal quark currents induced by the one-gluon exchange in
presence of the vacuum field and to formulation of the bosonization
procedure based on these nonlocal currents. A motivation and connection
of the model with QCD has been discussed as far as possible.
The present paper is concentrated on further development of the
model and on application to systematic calculations of the weak
decay constants and masses of mesons from different regions of the spectrum:
the light mesons and their excited states, heavy quarkonia and heavy-light
mesons.

With the minimal set of parameters (quark masses,
vacuum field strength and one quark-gluon coupling constant) the model
describes all qualitatively different regions of meson spectrum
to within ten percent inaccuracy. The reasons driving this successful
description can be easily recovered.

An interaction of quark spin with the vacuum gluon field,
contained in the quark propagator,
breaks the chiral symmetry and gives rise to the
splitting between masses of the pseudoscalar and vector mesons with
identical quark structure ($\rho-\pi$, $K-K^*$). This spin-field interaction
drives also the weak decay of pion and kaon.

Furthermore, the vacuum field produces three rigid
asymptotic regimes for the spectrum of collective modes.
The spectra of radial and orbital excitations of light
mesons are equidistant for $\ell\gg1$ or $n\gg1$, i.e., they have Regge
character. This is due to the specific form of
nonlocality of the quark and gluon propagators determined by confining
properties of the vacuum gluon field. After all, one concludes that
the confinement is responsible for Regge trajectories.
Localization of meson field at the center of masses
of a quark system provides other two asymptotic regimes.
In the limit of infinitely heavy quark, a mass of
quarkonium tends to be equal to a sum of the masses of constituent quarks,
while a mass of heavy-light meson approaches
the mass of a heavy quark: $M_{Q\bar Q}\to 2m_Q - \Delta_{Q\bar Q}$,
$M_{Q\bar q}\to m_Q + \Delta_{Q\bar q}$. The next-to-leading terms
$\Delta_{Q\bar Q}$ and $\Delta_{Q\bar q}$ do not depend on the heavy quark
mass. The same reasons provide the correct asymptotic behavior of
the weak decay constant for the heavy-light pseudoscalar mesons:
$f_P\sim 1/\sqrt{m_Q}$.

One can conclude that the
(anti-)self-dual homogeneous background gluon field determines
rather definitely a behavior of the masses and weak decay constants
in all different regions of the meson spectrum,
and this behavior is quantitatively consistent with experimental data.

Technically, these results are based on a decomposition of the
bilocal colorless quark currents into a series of nonlocal
currents with complete set of quantum numbers: spin, isospin,
radial and orbital numbers. Tensor structure of these nonlocal
currents is represented by the irreducible tensors of the
four-dimensional Euclidean rotational group, while their radial part
is determined by a specific form of gluon
propagator in the external vacuum field.
This decomposition provides a new point of view on the renormalization
problem: the Feynman diagrams appearing in each order of
perturbation theory are ultraviolet finite due to the nonlocality
of the meson-quark interaction.

The paper is organized as follows. In Sect.~II we review the main
points of the model with some modifications, that relate to choosing
the point of localization of meson field and description of super-fine
structure of the spectrum. These modifications reflect possibilities missed
in~\cite{efin}. In Sect.~III we consider the masses of light mesons
and their excitations, heavy quarkonia and heavy-light mesons, as well as
the weak decay constants.
The basic approximations of the model and problems for further
investigations are discussed in the last section.

\section{The model of induced nonlocal quark currents}
\subsection{Basic assumptions, approximations and notation}

The representation of the Euclidean generating functional for QCD,
in which the gluon and ghost fields are integrated out,
serves as a starting point for many models of hadronization.
Dyakonov and Petrov obtained this representation for the case of
nontrivial vacuum gluon field~\cite{dyakonov} (see also~\cite{efiv})
\begin{equation}
\label{gen2}
Z=\int d\sigma_{\rm vac}\int DqD\bar q
\exp\Biggl\{\int d^4x\sum_{f}^{N_F} \bar q_f(x)
\bigl(i\gamma_\mu\hat
\nabla_\mu-m_f\bigr)q_f(x)+\sum_{n=2}^{\infty}L_n\Biggr\} ,
\end{equation}
where $N_F$ is the number of flavors corresponding to the SU$(N_F)$
flavor group and
\begin{eqnarray}
&&L_n=\frac{g^n}{n!}\int d^4y_1...\int d^4y_n
j_{\mu _1}^{a_1}(y_1)\cdot ...
j_{\mu _n}^{a_n}(y_n)
G_{\mu _1...\mu_n}^{a_1...a_n}(y_1,...,y_n\mid B),
 \nonumber\\
&&j_\mu^a(y)=\sum_{f}^{N_F}\bar q_f(x)\gamma_\mu t^aq_f(x),
 \nonumber\\
&&~\hat\nabla_\mu=\partial_\mu-it^aB^a_\mu. \nonumber
\end{eqnarray}
The function $G_{\mu _1...\mu_n}^{a_1...a_n}$ is the exact
(up to the quark loops)
$n$-point gluon Green function in the external field $B_\mu ^a$.
We will investigate the mesonic $(q\bar q)$-collective modes
and consider Eq.~(\ref{gen2}) with the quark-quark interaction
truncated up to the term $L_2$
\begin{eqnarray}
\label{gen3}
Z=\int d\sigma_{\rm vac}\int DqD\bar q
\exp\Biggl\{ \int d^4x\sum_{f}^{N_F}\bar q_f(x)
\bigl(i\gamma_\mu\hat\nabla_\mu-m_f\bigr)q_f(x) \nonumber\\
+\frac{g^2}{2}\int\!\!\!\int d^4x{\rm d}^4y~j_\mu^a(x)
G_{\mu\nu}^{ab}(x,y\mid B)j_\nu^b(y)\Biggr\} .
\end{eqnarray}
Representations (\ref{gen2}) and (\ref{gen3}) imply, that there exists
some vacuum (classical) gluon field $B_\mu^a(x)$, which minimizes the
effective action (or effective potential) of the Euclidean QCD.
In the general case, the vacuum field depends on a set of parameters
$\{\sigma_{\rm vac}\}$, and the measure $d\sigma_{\rm vac}$ averages
all physical amplitudes over a subset of $\{\sigma_{\rm vac}\}$,
in respect to which the vacuum state is degenerate
(for more details see~\cite{efin} and references therein).

The quark-gluon interaction both in Eqs.~(\ref{gen2}) and (\ref{gen3})
are local, and a decomposition over degrees of $g^2$ generates a
renormalizable perturbation theory. It means, that an appropriate
regularization is implied in Eqs.~(\ref{gen2}) and (\ref{gen3}).
This point has to be stressed here,
since our final technical aim is a
transformation of the interaction term in Eq.~(\ref{gen3}),
which  generates completely new superrenormalizable perturbation expansion
of the functional integral~(\ref{gen3}).

Let us identify all ingredients of these general formulas for
the particular case of an homogeneous (anti-)self-dual vacuum field
\begin{eqnarray}
&& \hat{B}_\mu(x)=\hat{n}B_\mu(x),~~~~~~B_\mu(x)=B_{\mu\nu}x_\nu,
\nonumber\\
&& \hat{n}=n^at^a,~~~~~~~n^2=n^an^a=1. \nonumber
\end{eqnarray}
The constant tensor $B_{\mu\nu}$ satisfies the conditions:
\begin{eqnarray}
&&B_{\mu\nu}=-B_{\nu\mu},
~~B_{\mu\rho}B_{\rho\nu}=-B^2\delta_{\mu\nu}, \nonumber\\
&&\tilde B_{\mu\nu}=\frac{1}{2}
\varepsilon_{\mu\nu\alpha\beta}B_{\alpha\beta}=
\pm B_{\mu\nu},\nonumber
\end{eqnarray}
where $B$ is a gauge invariant tension of the vacuum field.
Since the chromomagnetic {\bf H} and chromoelectric {\bf E} fields 
relate to
each other like {\bf H}$=\pm${\bf E}, two spherical angles
$(\varphi,\theta)$ define a direction of the field in the Euclidean 
space.
In the diagonal representation of $\hat n=n^at^a$, an additional 
angle $\xi$
is needed to fix a direction of the field in the color space $$
\hat{n}=t^3\cos\xi+t^8\sin\xi~,~~~~~0\le\xi<2\pi.$$

The one-loop calculations and some nonperturbative estimations
of the effective potential for the homogeneous
gluon field argue (do not prove) that the potential could have a 
minimum at
nonzero value of the field tension $B\not=0$
(e.g., see~\cite{leutw1,leutw2,savvidy}).
We will assume, that the field under consideration realizes
a nonperturbative QCD vacuum, and study the manifestations of this field
in the spectrum of collective modes. Since
the effective potential is invariant under the Euclidean rotations,
parity and gauge transformations, this vacuum should be degenerate with
respect to the directions of the field in the color and Euclidean space 
and
should be the same for anti-self- and self-dual configurations.
According to this argumentation, the
field $B^a_\mu$ in (\ref{gen3}) corresponds to the tension $B$ minimizing 
the
effective potential, and the measure $d\sigma_{\rm vac}$
has the form
\begin{equation}
\label{sigma-vac}
\int d\sigma_{\rm vac}=\frac{1}{(4\pi)^2}
\int\limits_{0}^{\pi}{\rm d}\theta \sin\theta
\int\limits_{0}^{2\pi}{\rm d}\varphi
\int\limits_{0}^{2\pi}{\rm d}\xi\sum_{\pm}~,
\end{equation}
where the sign $\sum_{\pm}$ denotes averaging over the self- and
anti-self-dual configurations.
To simplify calculations
and to clarify the technical side of bosonization procedure
in presence of the background field,
we will omit the integral over $\xi$ in (\ref{sigma-vac}) and
fix a particular vector $n^a=\delta^{a8}$. In the fundamental
(matrix $t^8$) and adjoint (matrix $C^8$)
representations of $SU_{\rm c}(3)$ one gets
\begin{eqnarray}
&&\hat n=t^8=\mbox{diag}\Bigl(\frac{1}{\sqrt{3}},\frac{1}{\sqrt{3}},
-\frac{2}{\sqrt{3}}\Bigr)~,~~
\hat B_{\mu\rho}\hat B_{\rho\nu}=-(t^8)^2
B^2\delta_{\mu\nu},\nonumber\\
&&\breve n=C^8=\frac{\sqrt{3}}{2} K~,~~
\breve B_{\mu\rho}\breve B_{\rho\nu}=-\frac{3}{4}K^2B^2\delta_{\mu\nu}~,
\nonumber\\
&&K_{54}=-K_{45}=K_{76}=-K_{67}=i~,~K^2=\mbox{diag}(0,0,0,1,1,1,1,0).
\nonumber
\end{eqnarray}
The rest of elements of the matrix $K$ are equal to zero. It is 
convenient
to define the mass scale $\Lambda^2=\sqrt{3}B$:
\begin{eqnarray}
&&\breve B_{\mu\rho}\breve B_{\rho\nu}=
-\frac{1}{4}K^2\Lambda^4\delta_{\mu\nu},\nonumber\\
&&\hat B_{\mu\rho}\hat B_{\rho\nu}=
-v^2\Lambda^4\delta_{\mu\nu},~~~
v=\mbox{diag}\left(\frac{1}{3},\frac{1}{3},\frac{2}{3}\right).
\nonumber
\end{eqnarray}
We would like to stress, that the averaging over directions of the
background field in the color space should be incorporated into
the formalism, and its role should be analyzed.
But first of all, we would like to go as far as possible
with the formalism, that is simpler as possible, and test, how the
technique proposed in~\cite{efin} works in the meson phenomenology.

\subsection{Quark and gluon propagators}

The quark propagator $S_f(x,y\mid B)$ in Eq.~(\ref{gen3}) satisfies
the equation:
$$
(i\gamma_\mu\hat \nabla_\mu-m_f)S_f(x,y\mid B)=-\delta(x-y),
$$
and can be written in the form
\begin{eqnarray}
\label{qphase}
&&S_f(x,y\mid B)=e^{\frac{i}{2}(x\hat By)}
H_f(x-y\mid B)e^{\frac{i}{2}(x\hat By)}, \nonumber\\
&&H_f(z)=\frac{i\hat\nabla_\mu(z)\gamma_\mu+m_f}
{{\hat\nabla}^2(z)+m_f^2+(\sigma\hat B)}\delta(z),\\
\label{qsol}
&&\tilde H_f(p\mid B)=\frac{1}{2v\Lambda}\int \limits_{0}^{1}dt
e^{-\frac{p^2}{2v\Lambda^2}t}\left(\frac{1-t}{1+t}\right)
^{\frac{\alpha_f^2}{4v}}\left[\alpha_f+\frac{1}{\Lambda}p_\mu\gamma_\mu+
it\frac{1}{\Lambda}(\gamma fp)\right] \nonumber\\
&&~~~~~~~~~~~~~\times\left[P_\pm +P_\mp
\frac{1+t^2}{1-t^2}-\frac{i}{2}(\gamma f\gamma)\frac{t}{1-t^2}\right]~,
\end{eqnarray}
where
\begin{eqnarray}
&&P_{\pm}=\frac{1}{2}(1 \pm \gamma_5),
~~\alpha_f=\frac{m_f}{\Lambda}~,~~(xBy)=x_\mu B_{\mu\nu}y_\nu~,\nonumber\\
&&(pf\gamma)=p_\mu f_{\mu\nu}\gamma_\nu~,
~~f_{\mu\nu}=\frac{t^8}{v\Lambda^2}B_{\mu\nu}~,~~
f_{\mu\rho} f_{\rho\nu}=-\delta_{\mu\nu}~. \nonumber
\end{eqnarray}
The function $\tilde H_f$ is the Fourier transformed $H_f$.
The upper (lower) sign in the matrix $P_{\pm}$ corresponds to
the self-dual (anti-self-dual) field.

The term $(\sigma\hat B)=\sigma_{\alpha\beta}\hat B_{\alpha\beta}$
in Eq.~(\ref{qphase}) (the second line in Eq.~(\ref{qsol}))
describes an interaction of a quark spin with the background field.
One can see, that this spin-field interaction leads to the
singularity $1/m_f$ for $m_f\to 0$, which is a manifestation of the
zero mode (the lowest Landau level) of the massless Dirac equation
in the external (anti-)self-dual homogeneous field. The mathematical
point is that the spectrum of the operator $\gamma_\mu \partial_\mu$ is
continuous, whereas the spectrum of the operator
$\gamma_\mu\hat\nabla_\mu(x)$ is discrete and the lowest eigen
number is equal to zero.
Simple calculations give for $m_f\to 0$
\begin{eqnarray}
\label{chir0}
&&\tilde H_f(p\mid B)=2e^{-\frac{p^2}{2v\Lambda^2}}
\left[{1\over m_f}+{1\over m_f^2}(p_\mu\gamma_\mu+i(\gamma fp)\right]\cdot
\left[P_\mp-\frac{i}{4}(\gamma f\gamma)\right]+O(1)~,
\end{eqnarray}
and
\begin{eqnarray}
\label{chir1}
&&\lim_{\varepsilon\to 0} \lim_{m_f\to 0}
m_f\langle \bar q_f(x)q_f(x+\varepsilon)\rangle_B
=-\lim_{\varepsilon\to 0}\lim_{m_f\to 0}m_f{\rm Tr}
H_f(\varepsilon|B)\nonumber\\
&& =-\int \frac{d^4p}{(2\pi)^4}\lim_{m_f\to 0} m_f{\rm Tr}
\tilde H_f(p\mid B)=-\frac{4}{3\pi^2}\Lambda^4.
\end{eqnarray}
Due to the spin-field interaction the quark condensate is nonzero in
the limit of vanishing quark mass. This indicates that the chiral
symmetry is broken by the vacuum field in the limit $m_f\to 0$
(see also~\cite{leutw1}).
It will be clear below that just the spin-field interaction gives rise
to the splitting between the masses of the pseudoscalar and vector
mesons and provides a smallness of pion mass.

In terms of the variable $\zeta=p_\mu\gamma_\mu$ the propagator
$\tilde H_f(\zeta\mid B)$ is an entire analytical function in the complex
$\zeta$-plane. There are no poles corresponding to
the free quarks, which is treated as the confinement of quarks.
The following asymptotic behavior takes place:
\begin{equation}
\label{asymp}
\tilde H_f(\zeta\mid B)\rightarrow \left\{
\begin{array}{ll}
\frac{m_f+\zeta}{-\zeta^2}=
\frac{m_f+\gamma_\nu p_\nu}{p^2}
~\mbox{if}~
\zeta\rightarrow\pm \infty \
(p^2\rightarrow\infty) \\
O\bigl(\exp\bigl(\frac{\zeta^2}{2v\Lambda^2}\bigr)\bigr)=
O\bigl(\exp\bigl(\frac{-p^2}{2v\Lambda^2}\bigr)\bigr)
~\mbox{if}~
\zeta\rightarrow\pm i\infty \ (p^2\rightarrow -\infty).
\end{array}\right.
\end{equation}
Equations (\ref{asymp}) shows the standard local
behavior of the fermion propagator in the Euclidean region 
($p^2\to\infty$),
while in the physical region ($p^2\to -\infty$) we see
the exponential increase  typical for nonlocal theories
(for more details about the general theory of nonlocal interactions
of quantized fields see~\cite{efiv,mono}).
Below, the absence of the poles and the exponential
increase will be referred as the confinement properties of a propagator.

Function $D_{\mu\nu}^{ab}(x,y|B)$
in representation (\ref{gen3}) is the exact gluon propagator
for the pure gluodynamics  in presence of the vacuum field $B_\mu^a$.
This function is unknown, and some approximation has to be introduced.
For instance, the local  NJL model corresponds to the choice
$D_{\mu\nu}^{ab}=\delta^{ab}\delta_{\mu\nu}\delta(x-y)$.
We go beyond this approximation and replace the function
$D_{\mu\nu}^{ab}(x,y|B)$  by the confined part
\begin{eqnarray}
\label{confg}
&&D_{\mu\nu}(x,y\mid B)=
\delta_{\mu\nu}K^2
e^{\frac{i}{2}(x\breve By)}
D(x-y\mid \Lambda^2)
e^{\frac{i}{2}(x\breve By)},\\
&&D(z\mid \Lambda^2)=
\frac{1}{(2\pi)^2z^2}
\exp\Bigl\{-\frac{\Lambda^2z^2}{4}\Bigr\}\nonumber
\end{eqnarray}
of the gluon propagator
\begin{eqnarray}
\label{gphase}
G_{\mu\nu}^{ab}(x,y\mid B)=
D^{ab}_{\mu\nu}(x,y\mid B)+
R^{ab}_{\mu\nu}(x,y\mid B), \nonumber
\end{eqnarray}
which is a solution of the equation (for details see~\cite{efin})
\begin{eqnarray}
(\breve \nabla^2\delta_{\mu\nu}+
4i\breve B_{\mu\nu})G_{\nu\rho}(x,y\mid B)
=-\delta_{\mu\rho}\delta(x-y).\nonumber
\end{eqnarray}
The Fourier transform of the function $D(z\mid \Lambda^2)$ is an entire
analytical function in the momentum space. It has the local behavior
in the Euclidean region, but increases exponentially in the physical 
region.
This function describes a propagation of the confined modes of the
gluon field. Other terms $R^{ab}_{\mu\nu}$, that contain a contribution
of the zero modes and an anti-symmetric part, will be omitted.

Thus, the central point of our extension of the NJL-model consists in
taking into account the confining influence of the background field both
on the quark and gluon propagators.

\subsection{Color singlet bilocal quark currents}

Substituting gluon propagator (\ref{gphase}) to
the interaction term in representation~(\ref{gen3}),
using the Fierz transformation of the color,
flavor and Dirac matrices, and keeping only
the scalar $J^{aS}$, pseudoscalar $J^{aP}$, vector $J^{aV}$
and axial-vector $J^{aA}$ colorless currents,
we arrive at the expression~\cite{efin}
\begin{eqnarray}
\label{l23}
&&L_2=\frac{g^2}{2}\sum_{bJ}C_J
\int\!\!\!\int {\rm d}^4x{\rm d}^4y
J^{bJ}(x,y)D(x-y\mid \Lambda^2)J^{bJ}(y,x),\\
\label{cur1}
&&J^{bJ}(x,y)=\bar q_f(x)M^b_{ff^\prime}\Gamma^{J}e^{i(x\hat By)}
q_{f^\prime}(y)~, \\
&&\Gamma^S={\bf 1}~,~~\Gamma^P=i\gamma_5~,~~\Gamma^V=\gamma_\mu~,~~
\Gamma^A=\gamma_5\gamma_\mu~, \nonumber\\
&&C_S=C_P=\frac{1}{9}~,~~C_V=C_A
=\frac{1}{18}~.
\nonumber
\end{eqnarray}
Here $M^b_{ff^\prime}$ are the flavor mixing matrices 
($b=0,...,N_F^2-1$)
corresponding to the SU$(N_F)$ flavor group.
In the case of SU(2) and SU(3) they are given
by the matrices $\tau^b$ and $\lambda^b$ respectively.

Due to the phase factor $\exp\{i(x\hat By)\}$, bilocal quark currents
(\ref{cur1}) are the scalars under the local gauge transformations
\begin{eqnarray}
&&q(x)\rightarrow {\rm e}^{-i\hat \omega (x)}q(x)~,
~~\bar q(x)\rightarrow\bar q(x){\rm e}^{i\hat \omega (x)}~, \nonumber\\
&&\hat B_\mu \rightarrow {\rm e}^{-i\hat \omega (x)}
 \hat B_\mu {\rm e}^{i\hat \omega(x)} + \frac{i}{g}{\rm e}^{-i\hat
\omega (x)}\partial_\mu {\rm e}^{i\hat \omega (x)}.
\end{eqnarray}
Let us transform integration variables  $x$ and $y$
in Eq.~(\ref{l23}) to the coordinate system corresponding to the
center of masses of quarks $q_f(x)$ and $q_{f^\prime}(y)$
\begin{eqnarray}
\label{center}
x\rightarrow x+\xi_fy,~y\rightarrow x-\xi_{f^\prime}y,
~\xi_f=\frac{m_f}{m_f+m_{f^\prime}},~\xi_{f^\prime}
=\frac{m_{f^\prime}}{m_f+m_{f^\prime}}.
\end{eqnarray}
Corresponding transformation of the quark currents looks as
\begin{eqnarray}
\label{nccur}
&&J^{bJ}(x,y)=
\bar q_f(x)M^b_{ff^\prime}\Gamma^{J}e^{i(x\hat By)}q_{f^\prime}(y)
\rightarrow\nonumber\\
&&\bar q_f(x+\xi_fy)M^b_{ff^\prime}\Gamma^{J}e^{i(x\hat By)}
q_{f^\prime}(x-\xi_{f^\prime}y)=\nonumber\\
&&\bar q_f(x)M^b_{ff^\prime}\Gamma^Je^{\frac{1}{2}y
\stackrel{\leftrightarrow}{\nabla}_{ff^\prime}(x)}q_{f^\prime}(x)
\stackrel{{\rm ds}}{=}J^{bJ}(x,y),
\end{eqnarray}
where $\stackrel{\leftrightarrow}{\nabla}_{ff^\prime}$ is a linear
combination of the left and right covariant derivatives
\begin{eqnarray}
\stackrel{\leftrightarrow}{\nabla}_{ff^\prime}(x)=
\xi_{f}
(\stackrel{\leftarrow}{\partial}+i\hat B(x))
-\xi_{f^\prime}(\stackrel{\rightarrow}{\partial}-i\hat B(x)).
\nonumber
\end{eqnarray}
These covariant derivatives indicate, that the currents (\ref{nccur})
are nonlocal and colorless. Interaction term (\ref{l23}) takes the form
\begin{eqnarray}
\label{l31}
&&L_2=\frac{g^2}{8\pi^2}\sum_{bJ}C_J\int\!\!\!\int d^4xd^4y\frac{1}{y^2}
\exp\left\{-\frac{\Lambda^2y^2}{4}\right\}J^{bJ}(x,y)J^{bJ}(y,x),
\end{eqnarray}
where we have made use of the representation (\ref{confg}).
The currents are defined by Eq.~(\ref{nccur}). Transformation
(\ref{center}) turns out to be crucial for simultaneous description 
of the
light-light, heavy-light and heavy-heavy mesons.

\subsection{Decomposition of bilocal currents}
\label{dec}

An idea of our next step consists in a decomposition of the bilocal 
currents
(\ref{nccur}) over some complete set of orthonormalized polynomials in
such a way, that the relative coordinate of two quarks $y$ in
Eq.~(\ref{l31}) would be integrated out.
One can see, that a particular form of this set is determined
by the form of the gluon propagator
($\exp\{-\Lambda^2y^2/4\}$ in Eq.~(\ref{l31})).
The propagator plays the role of a weight function in the orthogonality
condition.
The physical meaning of the decomposition consists in classifying
a relative motion of two quarks in the bilocal currents
over a set of radial $n$ and angular $\ell$ quantum numbers.
In other words, according to general principles of quantum mechanics
the bilocal currents have to be represented as
a set of quark currents with definite radial $n$ and angular $\ell$
quantum numbers.
Thus, we are looking for a decomposition of the form
\begin{eqnarray}
\label{decomposition}
&&J^{bJ}(x,y)=\sum_{n\ell}(y^2)^{\ell/2}
f_{\mu_1...\mu_\ell}^{n\ell}(y)
{\cal J}_{\mu_1...\mu_\ell}^{bJ\ell n}(x)~,\\
&&f_{\mu_1...\mu_\ell}^{\ell n}(y)=L_{n\ell}(y^2)
T^{(\ell)}_{\mu_1...\mu_\ell}(n_y),~
n_y=y/\sqrt{y^2}~.\nonumber
\end{eqnarray}
The angular part of $f^{\ell n}$ is given by the irreducible tensors of 
the
four-dimensional rotational group $T^{(\ell)}_{\mu_1...\mu_\ell}$,
which are orthogonal
\begin{equation}
\label{angint}
\int_{\Omega} \frac{d\omega}{2\pi^2}
T^{(\ell)}_{\mu_1... \mu_\ell}(n_y)T^{(k)}_{\nu_1... \nu_k}(n_y)=
\frac{1}{2^\ell(\ell+1)}\delta^{\ell k}
\delta_{\mu_1\nu_1}...\delta_{\mu_\ell\nu_\ell}~,
\end{equation}
and satisfy the conditions:
\begin{eqnarray}
\label{gegen}
&& T^{(\ell)}_{\mu_1...\mu...\nu...\mu_\ell}(n_y)=
T^{(\ell)}_{\mu_1...\nu...\mu...\mu_\ell}(n_y)~,~~~~
T^{(\ell)}_{\mu...\mu...\mu_\ell}(n_y)=0, \nonumber\\
&& T^{(\ell)}_{\mu_1... \mu_\ell}(n_y)
T^{(\ell)}_{\mu_1... \mu_\ell}(n_{y^\prime})=
\frac{1}{2^\ell}C_\ell^{(1)}(n_yn_{y^\prime}).
\end{eqnarray}
The measure $d\omega$ in Eq.~(\ref{angint}) relates to integration over
the angles of unit vector $n_y$, and
$C_\ell^{(1)}$ in Eq.~(\ref{gegen}) are the Gegenbauer's (ultraspherical)
polynomials.
The polynomials
$L_{n\ell}(u)$ obey the condition:
\begin{displaymath}
\int \limits_{0}^{\infty}{\rm d}u\rho_\ell(u)
L_{n\ell}(u)L_{n^\prime\ell}(u)=\delta_{nn^\prime}~.
\end{displaymath}
The weight function $\rho_\ell(u)$ arising from the exponential
term in (\ref{l31}) looks like
$$
\rho_\ell(u)=u^\ell {\rm e}^{-u},$$
hence $L_{n\ell}(u)$ are the generalized Laguerre's polynomials.

The details of calculation of the currents
${\cal J}_{\mu_1...\mu_\ell}^{bJ\ell n}(x)$ in Eq.~(\ref{decomposition})
can be found in paper~\cite{efin}.
As a result, the interaction term $L_2$ takes the form
\begin{eqnarray}
&&L_2={1\over2}\sum_{bJ\ell n}\left({G_{J\ell n}\over\Lambda}\right)^2
\int {\rm d}^4x\Bigl[{\cal J}_{\mu_1...\mu_\ell}^{bJ\ell n}(x)\Bigr]^2,
\nonumber\\
\label{couplconst}
&&G_{J\ell n}^2=C_J~g^2~
\frac{(\ell+1)}{2^{\ell}n!(\ell +n)!}, \\
\label{cur3}
&&{\cal J}_{\mu_1...\mu_\ell}^{bJ\ell n}(x)=
\bar q(x)V_{\mu_1...\mu_\ell}^{bJ\ell n}(x)q(x),\\
\label{vertex2}
&&V_{\mu_1...\mu_\ell}^{bJ\ell n}(x)\equiv
V_{\mu_1...\mu_\ell}^{bJ\ell n}
\Biggl(\frac{\stackrel{\leftrightarrow}{\nabla}(x)}{\Lambda}\Biggr)
\nonumber\\
&&=M^b\Gamma^J\Bigl\{\!\!\Bigl\{F_{n\ell}
\Biggl(\frac{\stackrel{\leftrightarrow}{\nabla}^2(x)}{\Lambda^2}\Biggr)
T_{\mu_1...\mu_\ell}^{(\ell)}\Biggl(\frac{1}{i}
\frac{\stackrel{\leftrightarrow}{\nabla}(x)}{\Lambda}\Biggr)
\Bigr\}\!\!\Bigr\},\\
\label{ffactor}
&& F_{n\ell}(4s)=s^n~\int_0^1 dtt^{\ell+n}e^{st}.
\end{eqnarray}
The doubled brackets in
Eq.~(\ref{vertex2}) mean that the covariant derivatives commute
inside these brackets.
Form-factors $F_{n\ell}(s)$
are entire
analytical functions in the complex $s$-plane, which is a manifestation
of the gluon confinement.

The classification of the currents will be complete if we will
decompose ${\cal J}^{aJ\ell n}_{\alpha,\mu_1...\mu_\ell}$ with $J=V,A$
and $\ell>0$ into a sum of orthogonal currents
${\cal I}^{bJ\ell nj}_{\alpha\mu_1...\mu_\ell}$
with the different total angular momentum $j=\ell-1,\ell,\ell+1$.
Index $\alpha$ relates to the matrices $\Gamma^V_\alpha=\gamma_\alpha$ and
$\Gamma^A_\alpha=\gamma_5\gamma_\alpha$ in Eq.~(\ref{vertex2}).
This decomposition can be arranged by the following division
\begin{equation}
\label{j1}
{\cal J}^{bJ\ell n}_{\alpha,\mu_1...\mu_\ell}=
\sum_{j=\ell-1}^{\ell+1}{\cal I}^{bJ\ell nj}_{\alpha\mu_1...\mu_\ell},
\end{equation}
where
\begin{equation}
\label{i1}
{\cal I}^{bJ\ell nj}_{\alpha\mu_1...\mu_\ell}= \left\{
\begin{array}{lll}
{1\over(\ell+1)^2}{\cal P}_{\alpha\mu_1...\mu_\ell}\left[
\delta_{\alpha\mu_1}{\cal J}_{\rho,\rho\mu_2...\mu_\ell}^{bJ\ell n}
\right],
~j=\ell-1,\\
{1\over\ell+1}\sum_{i=1}^\ell\left[
{\cal J}_{\alpha,\mu_i...\mu_{i-1}\mu_{i+1}...\mu_\ell}^{bJ\ell n}-
{\cal J}_{\mu_i,\alpha...\mu_{i-1}\mu_{i+1}...\mu_\ell}^{bJ\ell n}\right]
,~j=\ell,\\
{1\over\ell+1}{\cal P}_{\alpha\mu_1...\mu_\ell}\left[
{\cal J}_{\alpha,\mu_1...\mu_\ell}^{bJ\ell n}-
{1\over\ell+1}
\delta_{\alpha\mu_1}{\cal J}_{\rho,\rho\mu_2...\mu_\ell}^{bJ\ell n}
\right],
~j=\ell+1.
\end{array}\right.
\end{equation}
Symbol ${\cal P}_{\alpha\mu_1...\mu_\ell}$ in (\ref{i1}) denotes a 
cyclic
permutation of the indices $(\alpha\mu_1...\mu_\ell)$.
Let $s_J$ be defined as
$$
s_{P}=s_{S}=0, \ \ s_{V}=s_{A}=1,
$$
then, using the orthogonality of the currents with different $j$,
we can rewrite interaction term $L_2$ as
\begin{eqnarray}
&&L_2=\sum_{aJ\ell n}\sum_{j=|\ell-s_J|}^{\ell+s_J}
\frac{1}{2\Lambda^2}G_{J\ell n}^2
\int d^4x\left[{\cal I}_{j}^{aJ\ell n}(x)\right]^2,
\nonumber
\end{eqnarray}
where we have introduced the notation
\begin{eqnarray}
&&{\cal I}_{\alpha}^{bJ0n1}={\cal J}_{\alpha}^{bJ0n}~~{\rm for}
~~J=V,A,~\ell=0,
\nonumber\\
&&{\cal I}_{\mu_1...\mu_\ell}^{bJ\ell n\ell}=
{\cal J}_{\mu_1...\mu_\ell}^{bJ\ell n}~~{\rm for}~~J=S,P,~\ell\ge0.
\nonumber
\end{eqnarray}
This form is equivalent to Eq.~(\ref{l23}), but now the interaction
between quarks is expressed in terms of the nonlocal quark currents,
that are elementary currents of the system in the sense of the
classification over quantum numbers.

For large Euclidean momentum the vertices $\tilde V^{aJln}$ behave as
$1/(p^2)^{1+\ell /2}$.
Therefore, only the "bubble" diagrams, shown in Fig.~1, are divergent.
These divergencies can be removed by the counter-terms of the form
$-2{\cal I}(x){\rm Tr}VS$.

To avoid an unnecessary complication of notation, it is convenient to
introduce condensed index ${\cal N}$ enumerating the currents with
all different combinations of the quantum numbers $a$, $J$, $\ell$, $n$
and $j$.
The renormalized vacuum amplitude $Z$ takes the form
\begin{eqnarray}
\label{gen4}
&&Z=\int d\sigma_{\rm vac}\int DqD\bar q
\exp\left\{\int\!\!\int {\rm d}^4x{\rm d}^4y \bar q(x)S^{-1}(x,y|B)q(y)
\right.\nonumber\\
&&
\left.+\sum_{\cal N}\frac{1}{2\Lambda^2}G^2_{\cal N}\int{\rm d}^4x
\left[{\cal I}_{\cal N}(x)
-{\rm Tr}V_{\cal N}S\right]^2\right\}.
\end{eqnarray}

\subsection{Bosonization}

By means of the standard bosonization procedure
\cite{klevansky,hatsuda} applied to Eq.~(\ref{gen4}) the amplitude $Z$
can be represented in terms of the composite
meson fields $\Phi_{\cal N}$~\cite{efin}:
\begin{eqnarray}
\label{genlast}
Z&=&N\int \prod_{\cal N}D\Phi_{\cal N}
\exp\left\{\frac{1}{2}\int\!\!\int {\rm d}^4x{\rm d}^4y
\Phi_{\cal N}(x)\left[\left(
\Box-M_{\cal N}^2\right)\delta(x-y)\right.\right. \nonumber\\
&-&\left.\left.h_{\cal N}^2\Pi_{\cal N}^R(x-y)\right]
\Phi_{\cal N}(y)+I_{\rm int}\left[\Phi\right]\right\},
\end{eqnarray}
\begin{eqnarray}
I_{\rm int}&=&
-\frac{1}{2}\int d^4x_1\int d^4x_2 h_{\cal N}h_{\cal N^\prime}
\Phi_{\cal N}(x_1)\left[\Gamma_{\cal N N'}(x_1,x_2)-
\delta_{\cal NN'}\Pi_{\cal N}(x_1-x_2)\right]\Phi_{\cal N'}(x_2)
\nonumber\\
&-&\sum_{m=3}\frac{1}{m}\int d^4x_1...\int d^4x_m
\prod_{k=1}^mh_{{\cal N}_k}\Phi_{{\cal N}_k}(x_k)
\Gamma_{{\cal N}_1...{\cal N}_m}(x_1,...,x_m),\nonumber\\
\Gamma_{{\cal N}_1...{\cal N}_m}&=&\int d\sigma_{\rm vac}
{\rm Tr}\left\{V_{{\cal N}_1}(x_1)S(x_1,x_2\mid B)...
V_{{\cal N}_m}(x_m)S(x_m,x_1\mid B)\right\}.\nonumber
\end{eqnarray}
Meson masses $M_{\cal N}$ are defined by the equations
\begin{equation}
\label{gep1}
1+\left(G_{\cal N}\over\Lambda\right)^2
\tilde \Pi_{\cal N}(-M_{\cal N}^2)=0,
\end{equation}
where $\tilde\Pi_{\cal N}(-M_{\cal N}^2)$ is the diagonal part of
the two-point function
$\tilde\Gamma_{\cal NN'}$, which corresponds
to the diagram shown in Fig.~2.a.
The fields $\Phi_{\cal N}$ (${\cal N}=\{a,J,\ell,n,j\}$)  with
$j>0$ satisfy the on-shell condition
$$p_\mu\tilde\Phi^{...\mu...}_{\cal N}(p)=0,
~{\rm if}~p^2=-M^2_{\cal N},$$
which excludes all extra degrees of freedom of the field, so that
the numbers $\ell$ and $j$ can be treated as the O(3) orbital momentum
and total momentum, respectively~\cite{efin}.
The total momentum $j$ plays the role of an observable spin of the
state with a given  ${\cal N}=\{b,J,\ell,n,j\}$.

The constants
\begin{equation}
\label{hmqq}
h_{\cal N}=1/\sqrt{\tilde\Pi'_{\cal N}(-M_{\cal N}^2)}
\end{equation}
play the role of the effective coupling constants of the
meson-quark interaction.

The quark masses $m_f$, the scale $\Lambda$ (strength of the 
background
field) and the quark-gluon coupling constant $g$ are the
free parameters of the effective meson theory 
(\ref{genlast})-(\ref{hmqq}).

In the one-loop approximation, the interactions between mesons with
given quantum numbers ${\cal N}=\{b,J,\ell,n,j\}$
are described by the quark loops like the diagram in Fig.~2.b.
The structure of diagrams in Fig.~2 is the same as in the standard 
NJL-model,
but in our case the quarks propagate in the vacuum gluon field, and 
the
meson-quark vertices are nonlocal. Due to this nonlocality the quark 
loops
are ultraviolet finite. The whole diagram is averaged by integration
over the measure $d\sigma_{\rm vac}$.

Figure~3 illustrates the central idea of the method of induced nonlocal
currents, that has been realized in this section. An effective
four-quark interaction is represented as an infinite
series of interactions between the nonlocal quark currents
characterized by the complete set of quantum numbers $\{b,J,\ell,n,j\}$.
The form of the currents is induced by a particular form of gluon
propagator. This new representation of the four-quark interaction
generates an expansion of any amplitude into the series of the
partial amplitudes with a particular value of the quantum numbers.
Each partial amplitude is ultraviolet finite at any order of
expansion over degrees of the coupling constant $g$.
The composite meson fields in Eq.~(\ref{genlast})
are nothing else but the "elementary" collective excitations,
that are classified according to the complete set of quantum numbers
of the relativistic two-quark system.

It should be stressed, that the model (\ref{genlast}) satisfies all
demands of the general theory of nonlocal interactions of
quantum fields~\cite{mono}, which means that Eq.~(\ref{genlast})
defines a nonlocal, relativistic, unitary and ultraviolet finite
quark model of meson-meson interactions.

Now we would like to test, how this formalism works
in the meson phenomenology.

\section{Meson spectrum and weak decay constants}

Let us rewrite Eq.~(\ref{gep1}) in more detailed form
\begin{equation}
\label{gep2}
\Lambda^2+G_{J\ell n}^2
\tilde\Pi_{bJ\ell nj}(-M_{bJ\ell nj}^2;m_f,m_{f^\prime};\Lambda)=0.
\end{equation}
The function $\tilde\Pi_{bJ\ell nj}$ in Eq.(\ref{gep2}) is given by
the diagonal part in the momentum representation of the tensor
\begin{eqnarray}
&& \Pi^{bJ\ell nj}_{\mu_1...,\nu_1...}(x-y;m_f,m_{f^\prime};\Lambda)=
\nonumber\\
&&\int d\sigma_{\rm vac}
{\rm Tr}\left[V_{\mu_1...}^{bJ\ell nj}(x)S(x,y\mid B)
V_{\nu_1...}^{bJ\ell nj}(y)S(y,x\mid B)\right].
\end{eqnarray}
Relation (\ref{gep2}) is the master equation for meson masses.
The function $\tilde\Pi$ can be calculated using the representations
(\ref{qphase}) and (\ref{qsol}) for the quark propagator and
(\ref{vertex2}) for the vertices. The only
point, that requires a comment, is an averaging over the space 
directions of
the vacuum field. Actually we have to average the tensors 
$f_{\mu\nu}$,
$f_{\mu\nu}f_{\alpha\beta}$ and so on. The generating formula 
looks as
$$
\langle\exp\left(if_{\mu\nu}J_{\mu\nu}\right)\rangle=
\frac{1}{4\pi}\int_{0}^{2\pi}d\varphi\int_{0}^{\pi}d\theta\sin\theta
\exp\left(if_{\mu\nu}J_{\mu\nu}\right)=\frac{\sin
\sqrt{2\left(J_{\mu\nu}J_{\mu\nu}\pm\tilde J_{\mu\nu}J_{\mu\nu}\right)}}
{\sqrt{2\left(J_{\mu\nu}J_{\mu\nu}\pm\tilde J_{\mu\nu}J_{\mu\nu}\right)}},
$$
where $\varphi$, $\theta$ are the spherical angles, $J_{\mu\nu}$ is an
anti-symmetric tensor, $\tilde J_{\mu\nu}$ is a dual tensor and
$\pm$ corresponds to the self- and anti-self-dual vacuum field.
In particular this general representation gives:
\begin{eqnarray}
&&\langle f_{\mu\nu}\rangle=0, \nonumber\\
&&\langle f_{\mu\nu}f_{\alpha\beta}\rangle=\frac{1}{3}
\left(\delta_{\alpha\mu}\delta_{\beta\nu}-
\delta_{\alpha\nu}\delta_{\beta\mu}\pm\varepsilon_
{\alpha\beta\mu\nu}\right).
\end{eqnarray}

\subsection{Light pseudoscalar and vector mesons}

First of all, let us fit the free parameters of the model, taking
the masses of $\pi$, $\rho$, $K$ and $K^*$ mesons as the basic 
quantities.
Below, we will sometimes use a symbol of a given meson instead of the
corresponding set of quantum numbers (for example, $\pi$ instead
of $(3,P,0,0;0)$).

Four equations for the masses of the basic mesons can be written 
in the form
\begin{eqnarray}
\label{fit1}
&&2\tilde \Pi_{\pi}(-M_\pi^2;m_u,m_u;\Lambda)=
\tilde \Pi_{\rho}(-M_\rho^2;m_u,m_u;\Lambda),\\
\label{fit2}
&&2\tilde \Pi_{K}(-M_K^2;m_s,m_u;\Lambda)=
\tilde \Pi_{K^*}(-M_{K^*}^2;m_s,m_u;\Lambda),\\
\label{fit3}
&&2\tilde \Pi_{\pi}(-M_\pi^2;m_u,m_u;\Lambda)=
\tilde \Pi_{K}(-M_K^2;m_s,m_u;\Lambda),\\
\label{fit4}
&&g^2=-9\Lambda^2/\tilde \Pi_{\pi}(-M_{\pi}^2;m_u,m_u;\Lambda).
\end{eqnarray}
If $M_\pi$, $M_\rho$, $M_K$ and $M_{K^*}$ are taken to be equal to the
experimental values, then the masses $m_u$ and $m_s$ of the $u$ and $s$
quarks
as the functions of $\Lambda$ are defined by 
Eqs.~(\ref{fit1}),(\ref{fit2}).
Using $m_u(\Lambda)$ and $m_s(\Lambda)$ in (\ref{fit3}),
we find the value of $\Lambda$, that provides a simultaneous 
description
of the strange and nonstrange mesons.  An optimal value of the 
coupling
constant $g$ is calculated by means of Eq.~(\ref{fit4}).
By this way we arrive at the values:
\begin{equation}
\label{param}
m_u=198.28~{\rm MeV},~m_s=412.96~{\rm MeV},
~\Lambda=319.46~{\rm MeV},~g=9.96.
\end{equation}
Solution (\ref{param}) is unique.

It is well-known, that
there should be a special reason, which provides a small pion mass
and splits the masses of pseudoscalar and vector mesons.
Breaking of chiral symmetry due to the four-quark interaction
and two independent coupling constants
for pseudoscalar and vector mesons ($g_P\not= g_V$ instead of our
parameter $g$) play the role of such reason in the local NJL-model.
As has already been pointed out, an interaction of quark spin with
the vacuum field leads to the singular behavior of the
quark propagator in the massless limit and generates a non-zero quark
condensate, which indicates breaking of the chiral symmetry by the
vacuum gluon field.  Now let us illustrate, that in our case the same
spin-field interaction is responsible for small pion mass and for the
mass-splitting between P- and V-mesons.

Polarization function $\tilde\Pi_{J}$ ($\ell=0$, $n=0$, $J=P,V$)
can be represented in the form
\begin{eqnarray}
\label{polx1}
&&\tilde\Pi_{J}(-M^2;m_f,m_{f^\prime};\Lambda)=\nonumber\\
&&-\frac{\Lambda^2}{4\pi^2}{\rm Tr}_v
\int_0^1dt_1\int_0^1dt_2\int_0^1ds_1\int_0^1ds_2
\left(\frac{1-s_1}{1+s_1}\right)^{\frac{m_f^2}{4v\Lambda^2}}
\left(\frac{1-s_2}{1+s_2}\right)^
{\frac{m_{f^\prime}^2}{4v\Lambda^2}}\cdot
\nonumber\\
&& \left[\frac{M^2}{\Lambda^2}
\frac{F_1^{(J)}(t_1,t_2,s_1,s_2)}{\Phi_2^4(t_1,t_2,s_1,s_2)}+
\frac{m_fm_{f^\prime}}{\Lambda^2}
\frac{F_2^{(J)}(s_1,s_2)}{(1-s_1^2)(1-s_2^2)
\Phi_2^2(t_1,t_2,s_1,s_2)}+\right.
\nonumber\\
&& \left.\frac{2v(1-4v^2t_1t_2)F_3^{(J)}(s_1,s_2)}
{\Phi_2^3(t_1,t_2,s_1,s_2)}\right]
\exp\left\{\frac{M^2}{2v\Lambda^2}\Phi(t_1,t_2,s_1,s_2)\right\},
\end{eqnarray}
where
\begin{eqnarray}
\label{polx2}
&&\Phi=\frac{\Phi_1(t_1,t_2,s_1,s_2)}{\Phi_2(t_1,t_2,s_1,s_2)},\\
&&\Phi_1=2v(t_1+t_2)[s_1\xi_1^2+s_2\xi_2^2]
+s_1s_2[1+4v^2t_1t_2(\xi_1-\xi_2)^2],\nonumber\\
&&\Phi_2=2v(t_1+t_2)(1+s_1s_2)+(1+4v^2t_1t_2)(s_1+s_2),
\nonumber\\
&&F_1^{(P)}=(1+s_1s_2)[A_1A_2+4v^2(t_1-t_2)^2\xi_1\xi_2s_1s_2],
\nonumber\\
&&F_1^{(V)}=\frac{1}{3}[(3-s_1s_2)A_1A_2+
4v^2(t_1-t_2)^2\xi_1\xi_2s_1s_2(1-3s_1s_2)],
\nonumber\\
&&A_1=[1-4v^2t_1t_2(\xi_1-\xi_2)]s_1+2v(t_1+t_2)\xi_2,\nonumber\\
&&A_2=[1+4v^2t_1t_2(\xi_1-\xi_2)]s_2+2v(t_1+t_2)\xi_1,\nonumber\\
\label{f2}
&&F_2^{(P)}=(1+s_1s_2)^2,~~~~~~ F_2^{(V)}=1-s_1^2s_2^2,\\
&&F_3^{(P)}=2(1+s_1s_2),~~~~~~ F_3^{(V)}=1-s_1s_2. \nonumber
\end{eqnarray}
Equations~(\ref{polx1})-(\ref{f2}) shows, that singularity
$(1-s_1)^{-1}(1-s_2)^{-1}$, arising from the  spin-field interaction
(see the second line of Eq.~(\ref{qsol})), is accumulated
in the term of Eq.~(\ref{polx1}) proportional to the quark masses.
Other terms are free from this singularity, although the spin-field
interaction contribute to them either. This is due to the structure
of the trace of the Dirac matrices.

Let us compare a behavior of pion and $\rho$-meson polarization
functions in the limit
$$m_f=m_{f^\prime}=m_u\ll\Lambda.$$
Using the singularity of the integrand in Eq.~(\ref{polx1}) at
$s_1\to1$ and $s_2\to1$,
one can check, that the pion polarization function is singular
in this limit and behaves as $1/m_u^2$:
\begin{eqnarray}
\label{polx3}
&&\tilde\Pi_{\pi}(-M^2;m_u,m_u;\Lambda)=
\nonumber\\
&& ~~~~~~~-{\rm Tr}_v \frac{4v^2\Lambda^4}{\pi^2m_u^2}
\int_0^1\!\!\int_0^1\frac{dt_1dt_2}{\Phi_2^2(t_1,t_2,1,1)}
\exp\left\{\frac{M^2}{2v\Lambda^2}\Phi(t_1,t_2,1,1)\right\}+O(1).
\end{eqnarray}
On the contrary, the $\rho$-meson polarization is regular at $m_u=0$ and
looks like
\begin{eqnarray}
\label{polx4}
&& \tilde\Pi_{\rho}(-M^2;m_u,m_u;\Lambda)=\nonumber\\
&& -\frac{\Lambda^2}{4\pi^2}{\rm Tr}_v
\int_0^1dt_1\int_0^1dt_2\int_0^1ds_1\int_0^1ds_2\left\{
\exp\left\{\frac{M^2}{2v\Lambda^2}\Phi(t_1,t_2,s_1,s_2)\right\}\cdot
\right.\\
&& ~~~~~\left[\frac{M^2}{\Lambda^2}
\frac{F_1^{(V)}(t_1,t_2,s_1,s_2)}{\Phi_2^4(t_1,t_2,s_1,s_2)}+
\frac{2v(1-4v^2t_1t_2)F_3^{(V)}(s_1,s_2)}{\Phi_2^3(t_1,t_2,s_1,s_2)}
\right]+
\nonumber\\
&& ~~~~~\left.\exp\left\{\frac{M^2}{2v\Lambda^2}
\Phi(t_1,t_2,s_1,1)\right\}\frac{2v}{\Phi_2^2(t_1,t_2,s_1,1)}
\right\}+O(m_u).\nonumber
\end{eqnarray}
This difference appears owing to the form of factors $F_2^{(P)}$
and $F_2^{(V)}$ in Eq.~(\ref{f2}) and leads to the inequality
$$|\tilde\Pi_{\pi}(-M^2;m_u,m_u;\Lambda)|\gg
|\tilde\Pi_{\rho}(-M^2;m_u,m_u;\Lambda)|.$$
This relation shows that the masses of pion and $\rho$-meson, satisfying
Eq.~(\ref{gep2}), are strongly splitted and $M^2_\rho\gg M^2_\pi$ when
the quark mass goes to zero. A similar picture takes place for $K$
and $K^*$ mesons, but since the strange quark mass is not so small,
the effect is more smooth.

Above consideration is illustrated in Table~\ref{ll}. The
pion mass is much larger, and difference in the masses of
pseudoscalar and vector mesons is smaller, if the spin-field interaction
in the quark propagator is eliminated (compare the first and the last 
lines
in the table).

Thus, we can conclude, that the large splitting between the masses of P-
and V- mesons is explained in our case by the spin-field interaction.
This splitting is the reason, why Eqs.~(\ref{fit1}) have appropriate
solution~(\ref{param}).

It should be noted, that the scalar polarization function
$\tilde\Pi_{S}$ differs from the pseudoscalar $\tilde\Pi_{P}$ only
by the sign before $m_fm_{f^\prime}$ in Eq.~(\ref{polx3}).
Due to the above-mentioned singularity, the term proportional to
$m_fm_{f^\prime}$ is leading, and $\tilde\Pi_{S}$ is positive for a
wide range of parameter values. As a result, Eq.~(\ref{gep2})
has no real solutions for the case of scalar mesons.
It looks interesting, that the scalar $(q\bar q)$ bound states do not
appear due to the same spin-field interaction, that diminishes 
the pion mass
and provides nonzero quark condensate.

Consideration of the $SU_F(3)$ singlet and the eighth octet states
shows an ideal mixing both for vector and pseudoscalar mesons.
The masses of $\omega$ and $\phi$, calculated with the parameter
values (\ref{param}), are
$$ M_\omega=M_\rho=770~{\rm MeV},~~~~~~~M_\phi=1034~{\rm MeV},$$
which is in a good agreement with the experimental values. An ideal
mixing of the pseudoscalar states is not the case, that can provide
an appropriate description of $\eta$ and $\eta^\prime$ mesons.
It is well known, that the problem of $\eta$ and $\eta^\prime$ masses
can be solved by taking into account another Euclidean gluon
configuration - the instanton vacuum field~\cite{t'hooft}.
The instantons can be incorporated into our formalism without any
principal problems, and we hope to realize this idea in forthcoming
publications.

Now let us consider the weak decays of $\pi$ and $K$ mesons.
In the lowest approximation, an amplitude of the decay $P\to l\bar\nu$
is given by the diagram in Fig~2.c. The weak decay constant $f_P$
is defined by the standard formulas
\begin{eqnarray}
\label{fp1}
&& A_{P\to l\bar\nu}(k,k^\prime)=i\frac{G_F}{\sqrt{2}}
{\cal K}h_PF(k^2)\Phi_P(k)k_\mu
\bar l(k^\prime)(1-\gamma_5)\gamma_\mu\nu(k+k^\prime),\nonumber\\
&& f_P=h_PF(-M^2),
\end{eqnarray}
where the meson-quark coupling constant $h_P$ is calculated {\it via}
Eq.~(\ref{hmqq}), and ${\cal K}$ is the KKM matrix element 
corresponding to
a given meson. For arbitrary pseudoscalar meson the diagram in Fig.~2.c
gives the following expression for $f_P$
\begin{eqnarray}
\label{fp2}
&& f_P=h_P\frac{1}{4\pi^2}{\rm Tr}_v\int\!\!\int\!\!\int_0^1
\frac{dtds_1ds_2(1+s_1s_2)}{[2vt(1+s_1s_2)+s_1+s_2]^3}
\left(\frac{1-s_1}{1+s_1}\right)^{\frac{m_f^2}{4v\Lambda^2}}
\left(\frac{1-s_2}{1+s_2}\right)^{\frac{m_{f^\prime}^2}{4v\Lambda^2}}
\cdot \nonumber\\
&& \left[m_f\frac{s_1+2vt[1-\xi_1(1+s_1^2)]}{1-s_1^2}+
m_{f^\prime}\frac{s_2+2vt[1-\xi_2(1+s_2^2)]}{1-s_2^2}
\right]\exp\left\{\frac{M^2}{2v\Lambda^2}\Psi(t,s_1,s_2)\right\},
\nonumber\\
&&\Psi=\frac{s_1s_2+2vt(s_1\xi_1^2+s_2\xi_2^2)}{2vt(1+s_1s_2)+s_1+s_2}.
\end{eqnarray}
The singularity of the integrand of Eq.~(\ref{fp2}) at $s_1\to1$ and
$s_2\to1$ appears from above-mentioned spin-field interaction
in the quark propagator and plays the main role in regulating the
value of $f_P$ for the light mesons.
Calculation of pion and kaon decay constants by means of Eq.~(\ref{fp2})
with the values of parameters (\ref{param}) gives
$$ f_\pi=126~{\rm MeV},~~~~~~~f_K=145~{\rm MeV}.$$
Note, that the coupling constants $h_\pi$ and $h_K$ depend on 
the meson mass,
quark masses and parameter $\Lambda$ (see Eq.~(\ref{hmqq}) and
Table~\ref{ll}).

One could get a definite impression, that simultaneous description of
the masses of $\pi$, $K$, $\rho$, $K^*$, $\omega$ and $\phi$ mesons,
and quite accurate values of $f_\pi$ and $f_K$ are obtained mostly
due to the breakdown of chiral symmetry by the spin-field
interaction (see also Eq.~(\ref{chir1})).

In order to clarify a status of this impression, one needs to
investigate the chiral limit of the model.
Whether or not, and in what form the Goldberger-Treiman and
Oakes-Reiner-Treiman relations are fulfilled in this limit?
Naively, it seems that the chiral limit corresponds to the case
$m_u\ll\Lambda$. However, the situation is much more complicated.

Just for illustration of this statement, consider the pion mass
in the limit $m_u\ll\Lambda$.
Integrals in Eq.~(\ref{polx3}) can be evaluated,
and we get the following asymptotic form of equation~(\ref{gep2}) for
pion mass
\begin{equation}
\label{pion}
1-g^2\frac{16\Lambda^6}{9\pi^2m_u^2M_\pi^4}{\rm Tr}_v v^2
\left[\exp\left\{\frac{M_\pi^2}{8v\Lambda^2}\right\}-
\exp\left\{\frac{M_\pi^2}{8v(1+2v)\Lambda^2}\right\}\right]^2=0.
\end{equation}
For any values of $g$ there is a real positive or negative solution
$M_\pi^2$ to Eq.~(\ref{pion}). The pion mass is equal to zero,
if the values of $m_u$, $g$ and $\Lambda$ satisfy the relation
$$ \frac{m_u^2}{\Lambda^2}=\frac{g^2}{9\pi^2}
{\rm Tr}_v\frac{v^2}{(1+2v)^2}\approx\left({2g\over15\pi}\right)^2.$$
For a fixed value of $g$ and $m_u/\Lambda\to 0$, the solution $M_\pi$
to Eq.~(\ref{pion}) is purely imaginary and behaves as
$$
\frac{M_\pi}{\Lambda}=i\frac{4}{3}\sqrt{7\ln\frac{\Lambda}{m_u}}
+O\left[\frac{\ln\ln(\Lambda/m_u)}{\sqrt{\ln(\Lambda/m_u)}}\right],
$$
which has nothing physically reasonable interpretation, but just 
indicates,
that the limit $g={\rm const}$, and $m_u/\Lambda\to0$ is ill-defined.

>From our point of view, a correct transition to the chiral limit has to be
based on a simultaneous changing of $m_u$, $g$ and $\Lambda$ as functions
of an actual physical parameter like the temperature or particle density.
The quark masses, coupling constant and the vacuum field strength
as the functions of this parameter have to be extracted from
consideration of QCD dynamics at nonzero temperature and density.
Unfortunately, this is a quite complicate problem, and we leave it for
further investigations.

The main result of this subsection is very simple:
we have demonstrated by the explicit model calculations, that the 
spin-field
interaction contained in the quark propagator
(in presence of the homogeneous (anti-)self-dual vacuum gluon field)
can be responsible for observable masses of the light pseudoscalar and 
vector
mesons (by exception of $\eta$, $\eta^\prime$), and for the values of the
weak decay constants of pion and kaon. All numerical results are
given in Tables \ref{par} and \ref{ll}.

\subsection{Regge trajectories}

It has been shown in our previous paper~\cite{efin}, that
the spectrum of radial and orbital excitations of the light mesons
is asymptotically equidistant:
\begin{eqnarray}
\label{mn}
M^2_{aJ\ell n}=\frac{8}{3}\ln\left(\frac{5}{2}\right)
\cdot\Lambda^2\cdot n
~+O(\ln n)~,~{\rm for}~~n\gg \ell,\\
\label{ml}
M^2_{aJ\ell n}=\frac{4}{3}\ln5\cdot\Lambda^2\cdot \ell~+O(\ln \ell),
~{\rm for}~~\ell\gg n.
\end{eqnarray}
Technically this result is based on the exponential
behavior of the quark propagator (\ref{asymp}) and vertex function
$F_{n\ell}$ (\ref{ffactor}) in
the Minkowski region ($p^2/\Lambda^2\to-\infty$) like
\begin{eqnarray}
\label{asymp1}
\tilde H_f(p\mid B)\rightarrow
O\left(\exp\left\{\frac{|p^2|}{2v\Lambda^2}\right\}\right),~
F_{n\ell}\left(p^2\right)\rightarrow
O\left(\exp\left\{\frac{|p^2|}{4\Lambda^2}\right\}\right),
\end{eqnarray}
and on the specific dependence of the coupling constant $G_{J\ell n}$
(see Eq.~(\ref{couplconst})) on the orbital and radial quantum
numbers $\ell$, $n$,
arising from the decomposition of the bilocal quark currents over the
generalized Lagguerre polynomials, which is determined in its order 
by the
form of gluon propagator (\ref{confg}) (for details see
sect.~\ref{dec}). In general words, the Regge character
of the spectrum is determined in our model by the confining properties
of the vacuum field.

Numerical calculation of masses of the first orbital excitations of
$\pi$, $K$, $\rho$ and $K^*$
mesons by means of Eq.~(\ref{gep2}) with the parameters~(\ref{param})
gives the masses shown in Table~\ref{lle}. The super-fine structure of the
excited states of $\rho$ and $K^*$ mesons coming from classification
of currents over total momentum (Eq.~(\ref{j1})) is qualitatively correct.
Super-fine splitting of the levels with $\ell=1$ is not very large.

\subsection{Heavy quarkonia}

Exponential behavior of the quark propagator and vertices (\ref{asymp1})
is responsible for the following relation between the masses of heavy
quarkonia $M_{Q\bar Q}$ and heavy quark $m_Q$ in the leading
approximation~\cite{efin}:
$$
M_{Q\bar Q}=2m_Q, \ \ \ \ {\rm for} \ m_Q\gg\Lambda.
$$
Now let us calculate the next-to-leading term in the mass formula.
In other words, we have to solve Eq.~(\ref{gep2}) with the polarization
function $\tilde\Pi_{J}$ defined by Eq.~(\ref{polx1}) with
\begin{equation}
\label{dhh}
m_f=m_{f^\prime}=m_Q\gg\Lambda, \ \ \ \ M_{Q\bar Q}=2m_Q-\Delta_{Q\bar Q}
\end{equation}
in the next-to-leading approximation over $1/m_Q$. Since the masses of 
quarks
are equal to each other, we have (see Eq.~(\ref{center}))
$$
\xi_1=\xi_2=1/2,
$$
which means that the composite quarkonium field
$\Phi_{Q\bar Q}(x)$
is localized at the center of masses of two heavy quarks
(in the Euclidean four-dimensional space). It is convenient to
transform variables $s_1$ and $s_2$ in Eq.~(\ref{polx1}):
\begin{equation}
\label{trhh}
r_1=(s_1+s_2)/\sqrt{2}, \ \ \ r_2=(s_1-s_2)/\sqrt{2}.
\end{equation}
The term with $F_3^{(J)}$ in Eq.~(\ref{polx1}) does not contribute
to the leading and next-to-leading behavior of the integral and can be
omitted. After the transformation we arrive at the expression:
\begin{eqnarray}
\label{polhh1}
&&\tilde\Pi_{J}(-M^2;m_Q,m_Q;\Lambda)=
\nonumber\\
&&-\frac{m_Q^2}{4\pi^2}{\rm Tr}_v
\int\!\!\!\int_0^1dt_1dt_2
\left(
\int_0^{1/\sqrt{2}}dr_1\int_{-r_1}^{r_1}dr_2+
\int_{1/\sqrt{2}}^{\sqrt{2}}dr_1\int_{-\sqrt{2}+r_1}^{\sqrt{2}-r_1}dr_2
\right)
\cdot
\nonumber\\
&&
\left[
\frac{R_1^{(J)}(t_1,t_2,r_1,r_2)}{\varphi_2^4(t_1,t_2,r_1,r_2)}+
\frac{R_2^{(J)}(r_1,r_2)}{(2-(r_1-r_2)^2)(1-(r_1+r_2)^2)
\varphi_2^2(t_1,t_2,r_1,r_2)}
\right]\cdot
\nonumber\\
&& \ \ \ \ \ \ \ \ \ \ \ \ \ \ \ \ \ \ \ \ \ \ \ \ \ \ \ \ \ \ \ \ \ \ \
\exp\left\{\frac{M^2}{4v\Lambda^2}\varphi(t_1,t_2,r_1,r_2)\right\}
+O\left(\frac{\Lambda}{m_Q}\right),
\end{eqnarray}
where
\begin{eqnarray}
\label{polhh2}
&&\varphi=\frac{\varphi_1(t_1,t_2,r_1,r_2)}{\varphi_2(t_1,t_2,r_1,r_2)}
-\frac{m_Q^2}{M^2}\ln
\frac{(\sqrt{2}+r_1)^2-r_2^2}{(\sqrt{2}-r_1^2)^2-r_2^2},
\\
&&\varphi_1=\sqrt{2}v(t_1+t_2)r_1+r_1^2-r_2^2
\nonumber\\
&&\varphi_2=v(t_1+t_2)(2+r_1^2-r_2^2)+\sqrt{2}(1+4v^2t_1t_2)r_1,
\nonumber\\
&&R_1^{(P)}=
\frac{1}{4}(2+r_1^2-r_2^2)[A_1A_2+v^2(t_1-t_2)^2(r_1^2-r_2^2)],
\nonumber\\
&&R_1^{(V)}=\frac{1}{12}[(6-r_1^2+r_2^2)A_1A_2+
v^2(t_1-t_2)^2(r_1^2-r_2^2)(1-3r_1^2+3r_2^2)],
\nonumber\\
&&A_1=r_1-r_2+\sqrt{2}v(t_1+t_2), \ \
A_2=r_1+r_2+\sqrt{2}v(t_1+t_2),\nonumber\\
&&R_2^{(P)}=(2+r_1^2-r_2^2)^2,\ \ R_2^{(V)}=4-(r_1^2+r_2^2)^2.
\nonumber
\end{eqnarray}
An asymptotic value  of the integral over $r_2$ in the limit
$M\gg\Lambda$ can be evaluated by the Laplace method. One can check, that
the function $\varphi$ has a maximum at the point $r_2=0$ for any values
of $r_1$, $t_1$, $t_2$:
\begin{eqnarray}
\label{ir2}
&&\frac{\partial}{\partial r_2}\varphi(t_1,t_2,r_1,r_2)|_{r_2=0}=0,
\nonumber\\
&&\frac{\partial^2}{\partial r_2^2}\varphi(t_1,t_2,r_1,r_2)|_{r_2=0}=
-\frac{m_Q^2}{M^2}\frac{8\sqrt{2}r_1}{(2-r_1^2)^2}-
\frac{4v(t_1+t_2)+\sqrt{2}r_1(1-v^2(t_1-t_2)^2)}
{\varphi_2(t_1,t_2,r_1,0)}<0,
\nonumber\\
&&\varphi(t_1,t_2,r_1,0)=
\frac{r_1[r_1+2\sqrt{2}v(t_1+t_2)]}{\varphi_2(t_1,t_2,r_1,0)}-
\frac{2m_Q^2}{M^2}\ln\frac{\sqrt{2}+r_1}{\sqrt{2}-r_1},
\end{eqnarray}
which means that the leading terms can be obtained by evaluating the
Gaussian integral over $r_2$. Furthermore, one can see, that the
largest value of the function $\varphi(t_1,t_2,r_1,0)$
in the interval $r_1\in[0,\sqrt{2}]$ corresponds to $r_1=0$ for any
$t_1$, $t_2$, moreover:
\begin{eqnarray}
\label{ir1}
\frac{\partial}{\partial r_1}\varphi(t_1,t_2,r_1,0)|_{r_1=0}=-
\frac{1}{\sqrt{2}}\left(\frac{4m_Q^2}{M^2}-1\right)<0,
\end{eqnarray}
and, in the leading approximation, the integrand is reduced to
the exponential function in $r_1$. Using Eqs.~(\ref{ir2}) and 
(\ref{ir1}),
and taking into account Eq.~(\ref{dhh}),
one can integrate over $r_2$ and $r_1$
with the result:
\begin{equation}
\label{hh1}
\tilde\Pi_{J}(-M^2;m_Q,m_Q;\Lambda)|=
-\frac{3\Lambda^3}{4\pi\sqrt{\pi}\Delta_{Q\bar Q}}
\int\!\!\!\int_0^1\frac{dt_1dt_2}{\sqrt{t_1+t_2}}
+O\left(\frac{\Lambda}{m_Q}\right).
\end{equation}
Integrating over $t_1$ and $t_2$ in Eq.~(\ref{hh1}) and
substituting the result to Eq.~(\ref{gep2}),
one can find
\begin{equation}
\label{hh2}
\frac{\Delta^{(J)}_{Q\bar Q}}{\Lambda}=
\frac{2(\sqrt{2}-1)}{\pi\sqrt{\pi}}C_Jg^2
+O\left(\frac{\Lambda}{m_Q}\right),
\end{equation}
where $C_P=1/9$, $C_V=1/18$ (see Eq.~(\ref{couplconst})).
It should be stressed, that the difference in the constants
\begin{equation}
\label{hh4}
\Delta^{(P)}_{Q\bar Q}=2\Delta^{(V)}_{Q\bar Q}
\end{equation}
originates from the Fierz transformation of the Dirac matrices in
the interaction term $L_2$ in representation~(\ref{gen3}).
Relation~(\ref{hh4}) means that
the vector quarkonium state is always heavier than the 
pseudoscalar one.

The results of numerical calculation of the masses of different heavy
quarkonia states are summarized in Tables \ref{cc} and \ref{bb}.
The parameters $\Lambda$ and $g$ are equal to the values~(\ref{param})
fitting the light meson masses, and $m_c=1650$~MeV, $m_b=4840$~MeV.
An agreement
with the experimental values is rather satisfactory.
The super-fine splitting ($\chi_{c0}$, $\chi_{c1}$, $\chi_{c2}$ and so on)
is very small, since it is
 regulated by the terms $O(1/m_Q)$ in Eq.(\ref{gep2}). Its description
is qualitatively correct.
The splitting is generated in our model by
dividing the quark currents with $\gamma_\mu$ and $\ell>0$ into the
antisymmetric, symmetric traceless and diagonal parts (see Eq.~(\ref{j1})),
which extracts the states with different total angular momentum, mixed
in the currents
$\bar qM\gamma_\alpha T^{(\ell)}_{\mu_1...\mu_\ell}F_{n\ell}q$.

We conclude that the correct description of the
heavy quarkonia in our model is provided by the specific form of
nonlocality of the quark and gluon propagators induced by
the vacuum field, localization of meson field at
the center of masses of constituent quarks and by a separation of the
nonlocal currents with different total momentum.
In general, the spectrum is driven by the rigid asymptotic
formulas~(\ref{dhh}) and (\ref{hh2}).

\subsection{Heavy-light mesons}

Another interesting sector of meson spectrum is the heavy-light
mesons, characterized by a rich physics~\cite{iws,nuebert}.
In this subsection we will consider the masses
and weak decay constants of heavy-light mesons.
First of all, let us get the asymptotic formulas in the limit of 
infinitely
heavy quark. Namely, we have to investigate the behavior of
the polarization function $\tilde\Pi_{J}(-M;m_Q,m_q;\Lambda)$
(Eq.~\ref{polx1}) and the weak decay constant $f_P$ (Eq.~\ref{fp2})
in the case
\begin{eqnarray}
\label{hl1}
&&m_f=m_Q\gg\Lambda, \ \ \ \ m_{f^\prime}=m_q\sim O(\Lambda),
\nonumber\\
&&\xi_f=\frac{m_Q}{m_Q+m_q}=1+O(m_q/m_Q), \ \ \ \
\xi_{f^\prime}=\frac{m_q}{m_Q+m_q}=O(m_q/m_Q).
\end{eqnarray}
Equations~(\ref{hl1}) indicate, that in the heavy quark limit the 
composite
meson field $\Phi_{Q\bar q}(x)$ is localized at the point, in which the
heavy quark $Q$ is situated.

Let us show,
that in the limit~(\ref{hl1})
the leading and next-to-leading terms of the
solution to Eq.~(\ref{gep2}) read
\begin{equation}
\label{hl11}
M_{Q\bar q}=m_Q+\Delta^{(J)}_{Q\bar q}+O(1/m_Q),
\end{equation}
where the next-to-leading term $\Delta_{Q\bar q}^{(J)}$ does not
depend on the heavy quark mass
$m_Q$. This term is a function of a light quark mass
$m_q$ and coupling constant $G_{J00}$ (see Eq.~(\ref{couplconst})).

Omitting the term with $F_3^{(J)}$, which does not contribute
to the leading and next-to-leading
behavior of the integral
and taking into account conditions~(\ref{hl1}),
one can rewrite~Eq.(\ref{polx1}) in the form
\begin{eqnarray}
\label{hl2}
&&\tilde\Pi_{J}(-M^2;m_Q,m_q;\Lambda)=
-\frac{1}{4\pi^2}{\rm Tr}_v
\int\!\!\!\int\!\!\!\int\!\!\!\int_0^1dt_1dt_2ds_1ds_2
\left(\frac{1-s_2}{1+s_2}\right)^{\frac{m_q^2}{4v\Lambda^2}}
\cdot
\nonumber\\
&&\left[
\frac{(1-4v^2t_1t_2)Y(t_1,t_2,s_2)T_1^{(J)}(s_1,s_2)s_1M^2}
{[s_1X(t_1,t_2,s_2)+Y(t_1,t_2,s_2)]^4}+\right.
\nonumber\\
&&\left.
\frac{T_2^{(J)}(s_1,s_2)m_Qm_q}{(1-s_1^2)(1-s_2^2)
[s_1X(t_1,t_2,s_2)+Y(t_1,t_2,s_2)]^2}
\right]\cdot
\nonumber\\
&& \ \ \ \ \ \ \ \ \ \ \ \ \ \ \ \ \ \ \ \ \ \ \ \ \ \ \ \ \ \ \ \ \ \ \
\exp\left\{\frac{M^2}{2v\Lambda^2}\phi(t_1,t_2,s_1,s_2)\right\}
+
O\left(\frac{\Lambda}{m_Q}\right),
\end{eqnarray}
where
\begin{eqnarray}
&&\phi=\frac{s_1Y(t_1,t_2,s_2)}{s_1X(t_1,t_2,s_2)+Y(t_1,t_2,s_2)}
-\frac{m_Q^2}{2M^2}\ln\frac{1+s_1}{1-s_1},
\nonumber\\
&&X=1+4v^2t_1t_2+2v(t_1+t_2)s_2
\nonumber\\
&&Y=2v(t_1+t_2)+(1+4v^2t_1t_2)s_2
\nonumber\\
&&T_1^{(P)}=1+s_1s_2, \ \ \ \ T_1^{(V)}=\frac{1}{3}(3-s_1s_2),
\nonumber\\
&&T_2^{(P)}=(1+s_1s_2)^2,\ \ \ \ T_2^{(V)}=1-s_1^2s_2^2.
\nonumber
\end{eqnarray}
One can check, that for any $t_1$, $t_2$ and $s_2$
the function $\phi(t_1,t_2,s_1,s_2)$ has a maximum at
$s_1=s_1^{\rm max}$:
\begin{eqnarray}
\label{hl4}
&&s_1^{\rm max}=
\frac{Y(t_1,t_2,s_2)}{2X(t_1,t_2,s_2)}\left(1-\frac{m_Q^2}{M^2}\right)+
O\left(\frac{\Lambda^2}{m_Q^2}\right),\nonumber\\
&&\phi(t_1,t_2,s_1^{\rm max},s_2)=
\frac{Y(t_1,t_2,s_2)}{2X(t_1,t_2,s_2)}\left(1-\frac{m_Q^2}{M^2}\right)^2+
O\left(\frac{\Lambda^3}{m_Q^3}\right),\nonumber\\
&&\frac{\partial^2}{\partial s_1^2}
\phi(t_1,t_2,s_1,s_2)|_{s_1=s_1^{\rm max}}
=-\frac{2X(t_1,t_2,s_2)}{Y(t_1,t_2,s_2)}
\left(1-\frac{m_Q^2}{M^2}\right)^2+
O\left(\frac{\Lambda^3}{m_Q^3}\right).
\end{eqnarray}
Therefore, we can write
\begin{eqnarray}
\label{hl5}
&&\tilde\Pi_{J}(-M^2;m_Q,m_q;\Lambda)=\nonumber\\
&&-\frac{1}{4\pi^2}{\rm Tr}_v
\int\!\!\!\int\!\!\!\int_0^1dt_1dt_2ds_2
\left(\frac{1-s_2}{1+s_2}\right)^{\frac{m_q^2}{4v\Lambda^2}}
\exp\left\{\frac{M^2}{4v\Lambda^2}\left(1-\frac{m_Q^2}{M^2}\right)^2
\frac{Y(t_1t_2,s_2)}{X(t_1,t_2,s_2)}\right\}
\cdot
\nonumber\\
&& \ \ \ \ \ \ \ \ \ \ \ \ \ \ \ \ \ \ \ \ \ \ \ \
\left[
\frac{(1-4v^2t_1t_2)s^{\rm max}_1M^2}
{Y^3(t_1,t_2,s_2)}+
\frac{m_Qm_q}{(1-s_2^2)Y^2(t_1,t_2,s_2)}
\right]\cdot
\nonumber\\
&&\ \ \ \ \ \ \ \ \ \ \ \ \ \ \ \ \ \ \ \ \ \ \ \ \
\int_{-\infty}^{\infty}ds_1
\exp\left\{-\frac{M^2}{2v\Lambda^2Y(t_1,t_2,s_2)}s_1^2\right\}
+O\left(\frac{\Lambda}{m_Q}\right).
\end{eqnarray}
Integrating out the variable $s_1$ in Eq.~(\ref{hl5}),
substituting $M=m_Q+\Delta^{(J)}_{Q\bar q}$ to the resulting
expression  and using Eq.~(\ref{gep2}), we arrive at the equation
\begin{eqnarray}
\label{hl7}
&&1=
g^2C_J\frac{1}{(2\pi)^{3/2}}{\rm Tr}_v\sqrt{v}
\int\!\!\!\int\!\!\!\int_0^1\frac{dt_1dt_2ds_2
\left[(1-s_2)/(1+s_2)\right]^{\frac{m_q^2}{4v\Lambda^2}}}
{\left[X(t_1,t_2,s_2)Y(t_1,t_2,s_2)\right]^{3/2}}
\cdot
\nonumber\\
&&
\left[\frac{\Delta^{(J)}_{Q\bar q}}{\Lambda}+
\frac{X(t_1,t_2,s_2)}{1-s_2^2}\frac{m_q}{\Lambda}
\right]
\exp\left\{\frac{\left[\Delta^{(J)}_{Q\bar q}\right]^2}{v\Lambda^2}
\frac{Y(t_1t_2,s_2)}{X(t_1,t_2,s_2)}\right\}
+O\left(\frac{\Lambda}{m_Q}\right).
\end{eqnarray}
Equation~(\ref{hl7}) describes dependence of $\Delta^{(J)}_{Q\bar q}$
in mass formula~(\ref{hl11})
on the coupling constant $g$, the light quark mass $m_q$ and vacuum
field strength $B$ ($\Lambda$).  There is a single real solution to
Eq.~(\ref{hl7}) for any positive $g$, $m_q$ and $\Lambda$. In particular,
for the values (\ref{param}) we get
\begin{eqnarray}
\Delta^{(P)}_{Q\bar u}=20 \ {\rm MeV} , \ \ \ \
\Delta^{(V)}_{Q\bar u}=155 \ {\rm MeV},
\nonumber\\
\Delta^{(P)}_{Q\bar s}=63 \ {\rm MeV} , \ \ \ \
\Delta^{(V)}_{Q\bar s}=191 \ {\rm MeV}.
\nonumber
\end{eqnarray}
As is seen from Eq.~(\ref{hl7}), the difference between pseudoscalar
$\Delta^{(P)}_{Q\bar q}$ and vector $\Delta^{(V)}_{Q\bar q}$ is due to
the constant $C_J$, that appears from the Fierz transformation of 
the Dirac
matrices. This is the same situation as in the case of heavy quarkonia
(see Eq.~(\ref{hh2})).

Table \ref{hl} demonstrates reasonably good agreement between
the experimental data and the masses of the heavy-light mesons
calculated by means of Eq.~(\ref{gep2})
with the parameters~(\ref{param}).
The masses of heavy quarks are the same as in the description of
the heavy quarkonia (see Table~\ref{par}).

Now let us turn to the calculation of the weak decay constant for the
pseudoscalar heavy-light mesons. Under conditions~(\ref{hl1}), 
the integral
over $s_1$ in Eq.~(\ref{fp2}) can be evaluated by the Laplace method.
The result is
\begin{eqnarray}
\label{hlf1}
f_P=h_P\frac{\Lambda^2}{m_Q}
{\cal A}_f\left(\frac{\Delta^{(P)}_{Q\bar q}}{\Lambda},
\frac{m_q}{\Lambda}\right),
\end{eqnarray}
where
\begin{eqnarray}
\label{hlf2}
&&{\cal A}_f={\rm Tr}_v\frac{\sqrt{v}}{(2\pi)^{3/2}}
\int\!\!\int_0^1
\frac{dtds_2\left[(1-s_2)/(1+s_2)\right]^{\frac{m_q^2}{4v\Lambda^2}}}
{\left[s_2+2vt\right]^{3/2}\left[1+2vts_2\right]^{3/2}}\times
\nonumber\\
&&\left[\frac{\Delta^{(P)}_{Q\bar q}}{\Lambda}+\frac{m_q}{\Lambda}
\frac{1+2vts_2}{1-s_2^2}\right]
\exp\left\{\frac{\left[\Delta^{(P)}_{Q\bar q}\right]^2}{v\Lambda^2}
\frac{s_2+2vt}{1+2vts_2}\right\}
+O\left(\frac{\Lambda}{m_Q}\right),
\end{eqnarray}
and the difference $\Delta^{(P)}_{Q\bar q}$ between the masses of
heavy quark and heavy-light meson is given by Eq.~(\ref{hl7}).
The procedure for obtaining Eqs.~(\ref{hlf1}) and (\ref{hlf2}) is
very similar to the calculations providing Eq.~(\ref{hl7}).
To get the final formula for
$f_P$, the asymptotic form
of meson-quark coupling constant $h_P$ has to be defined
in the case~(\ref{hl1}).
Performing calculations analogous to that ones, which lead
to Eq.~(\ref{hl7}), we arrive at
\begin{eqnarray}
\label{hlf3}
&&h_P=\sqrt{\frac{m_Q}{\Lambda}}
{\cal A}^{-1}_h\left(\frac{\Delta^{(P)}_{Q\bar q}}{\Lambda},
\frac{m_q}{\Lambda}\right),\nonumber\\
&&{\cal A}^2_h=\frac{\Delta^{(P)}_{Q\bar q}}{2(2\pi)^{3/2}\Lambda}
{\rm Tr}_v\frac{1}{\sqrt{v}}
\int\!\!\!\int\!\!\!\int_0^1\frac{dt_1dt_2ds_2
\left[(1-s_2)/(1+s_2)\right]^{\frac{m_q^2}{4v\Lambda^2}}}
{\left[X(t_1,t_2,s_2)Y(t_1,t_2,s_2)\right]^{3/2}}
\cdot
\nonumber\\
&&
\left[\frac{\Delta^{(J)}_{Q\bar q}}{\Lambda}+
\frac{X(t_1,t_2,s_2)}{1-s_2^2}\frac{m_q}{\Lambda}
\right]
\exp\left\{\frac{\left[\Delta^{(J)}_{Q\bar q}\right]^2}{v\Lambda^2}
\frac{Y(t_1,t_2,s_2)}{X(t_1,t_2,s_2)}\right\}
+O\left(\frac{\Lambda}{m_Q}\right).
\end{eqnarray}
One can see, that
Eqs.~(\ref{hlf1})-(\ref{hlf3}) give the following
asymptotic relation in the heavy quark limit~(\ref{hl1})
\begin{equation}
\label{hlf4}
f_P=\frac{\Lambda^{3/2}}{\sqrt{m_Q}}\frac{{\cal A}_f}{{\cal A}_h},
\end{equation}
where ${\cal A}_f$ and ${\cal A}_h$ do not depend on the heavy 
quark mass
in the leading approximation over $\Lambda/m_Q$, as is indicated in
(\ref{hlf2}),(\ref{hlf3}).
Relation~(\ref{hlf4}) agrees with
the accepted notion about behavior of the weak decay
constants of the heavy-light mesons~\cite{nuebert}.
Results of numerical calculation of the weak decay constants for
different pseudoscalar mesons are given in Table~\ref{hl}.

\section{Discussion}

For conclusion, we would like to point out several problems, 
that require
more profound studying.

We have assumed from the very beginning, that
the nonperturbative QCD vacuum is characterized by a nonzero 
background
(anti-)self-dual homogeneous field. In other words, a
minimum of the QCD effective potential (the free energy density) for
this gluon configuration is assumed to be at nonzero field strength.
Different estimations of the effective potential indicates, that
this situation can be realized~\cite{savvidy}.
Although these estimations
cannot be used as a basis for more or less rigorous
proof, they underline the key role of the asymptotic freedom in
forming the effective potential for an homogeneous gluon field.
Just the asymptotic freedom is the distinctive feature of nonabelian
gauge theories like QCD.
Gluon self-interaction is manifested also in the nontrivial form of
gluon propagator (\ref{confg}).
Although the background field is quasi-abelian,
the nonabelian nature of the gluon field plays the crucial role for 
above
considered model.

In order to clarify the basic assumption of this paper,
one needs to get a reliable nonperturbative estimation of the
free energy density or effective potential of QCD for the background
field under consideration. Lattice calculations seem to be the most
promising approach to this problem.

In this paper, we have demonstrated how the singularity $1/m_f$ of
the quark propagator affects the masses and weak decay constants
of light mesons. However, more detailed consideration
of the chiral symmetry breaking by the background field is needed.
This can be achieved by investigating the Dirac
equation in presence of the homogeneous (anti-)self-dual field.
Another source of violation of the chiral symmetry is the effective
four-quark interaction. The divergent diagram in Fig.~1 should play
the key role in studying of this mechanism of the symmetry breaking.
One can expect that additional breakdown of chiral symmetry
by the four-quark interaction could diminish the
quark masses. They come into our formalism as the current masses,
but their fitted values (see Table~\ref{par})
are close to the constituent quark masses rather than to the current 
ones.

One could see, that the coupling constant $g$ in Table~\ref{par}
is rather large.
A possible origin of this unpleasant feature could be covered
in elimination of some terms of the gluon propagator (\ref{confg})
(for more details see~\cite{efin}). In other words, some
truncations in the gluon propagator was compensated by the rising
of the coupling constant. This point also has to be investigated 
carefully.

\vspace*{.5cm}

\begin{center}

{\bf Acknowledgments}

\end{center}

We would like to thank J.~H\"ufner, F.~Lenz and F.~Sch\"oberl for
interesting discussions. We are also grateful to our colleagues
A.~Dorokhov,  A.~Efremov, S.~Gerasimov, M.~Ivanov, N.~Kochelev,
V.~Liubovitsky,
S.~Mikhailov and O.~Teriaev for many stimulating questions they 
asked at
the seminars in the Bogoliubov Laboratory of Theoretical Physics.
One of the authors (S.N.N.) would like to thank H.~Leutwyler for
discussion and valuable comments.

This work was supported by the Russian Foundation for Basic Research
under grant No.~94-02-03463-a and by the Russian University Foundation
under grant No.~94-6.7-2042.

\newpage

\begin{table}
\caption{Parameters of the model.}
\label{par}

\begin{center}

\begin{tabular}{ccccccc}
\hline
\hline
$m_u$ (MeV)
&$m_d$ (MeV)&$m_s$ (MeV)&$m_c$ (MeV)&$m_b$ (MeV)&$\Lambda$ (MeV)&$g$  \\
\hline
198.3&198.3&413&1650&4840&319.5&9.96\\
\hline
\hline
\end{tabular}

\end{center}

\end{table}

\begin{table}
\caption{The masses(MeV), weak decay constants (MeV)
       and meson-quark coupling constants $h$ of the
       light mesons. $M^*$ - calculation without
       taking into account the spin-field interaction.}
\label{ll}

\begin{center}

\begin{tabular}{lcccccc}
\hline
\hline
{\small  Meson }
&$\pi$&$\rho$&$K$&$K^*$&$\omega$&$\phi$  \\
\hline
$M$&140&770&496&890&770&1034\\
$M^{\rm exp}$&140&770&496&890&786&1020\\
$f_P$&126&-&145&-&-&-\\
$f_P^{\rm exp}$&132&-&157&-&-&-\\
$h$&6.51&4.16&7.25&4.48&4.16&4.94\\
$M^*$&630&864&743&970&864&1087\\
\hline
\hline
\end{tabular}

\end{center}

\end{table}

\begin{table}
\caption{ The masses (MeV) of orbital excitations of $\pi$, $K$, $\rho$ and
       $K^*$ mesons. Super-fine structure of the $\ell=1$ excitation
       of $\rho$ and $K^*$ is shown ($\ell$ is the orbital momentum and
       $j$ is the total momentum (an observable spin) of a state).}
\label{lle}

\begin{center}

\begin{tabular}{lcccc}
\hline
\hline
{\small Meson}&$\ell$&$j$&$M$&$M^{\rm exp}$  \\
\hline
$\pi$&0&0&140&140\\
$b_1$&1&1&1252&1235\\
\hline
$K$&0&0&496&496\\
$K_1(1270)$&1&1&1263&1270\\
\hline
$\rho$&0&1&770&770\\
&1&0&1238&\\
$a_1$&1&1&1311&1260\\
$a_2$&1&2&1364&1320\\
\hline
$K^*$&0&1&890&890\\
&1&0&1274&\\
$K_1$(1400)&1&1&1342&1400\\
$K^*_2$&1&2&1388&1430\\
\hline
\hline
\end{tabular}

\end{center}

\end{table}

\begin{table}
\caption{The spectrum of charmonium.}
\label{cc}

\begin{center}

\begin{tabular}{lccccccc}
\hline
\hline
{\small Meson }& $\eta_c$&$J/\psi$&$\chi_{c_0}$&$\chi_{c_1}$&$\chi_{c_2}$
&$\psi^\prime$&$\psi^{\prime\prime}$\\
\hline
$n$&0&0&0&0&0&1&2\\
$\ell$&0&0&1&1&1&0&0\\
$j$&0&1&0&1&2&1&1\\
$M$ (MeV)&3000&3161&3452&3529&3531&3817&4120\\
$M^{\rm exp}$ (MeV)&2980&3096&3415&3510&3556&3770&4040\\
\hline
\hline
\end{tabular}

\end{center}

\end{table}

\begin{table}
\caption{The spectrum of bottomonium.}
\label{bb}

\begin{center}

\begin{tabular}{lccccccccc}
\hline
\hline
{\small Meson }& $\Upsilon$&$\chi_{b_0}$&$\chi_{b_1}$&$\chi_{b_2}$
&$\Upsilon^\prime$&$\chi_{b_0}^\prime$&$\chi_{b_1}^\prime$
&$\chi_{b_2}^\prime$&$\Upsilon^{\prime\prime}$\\
\hline
$n$&0&0&0&0&1&1&1&1&2\\
$\ell$&0&1&1&1&0&1&1&1&0\\
$j$&1&0&1&2&1&0&1&2&1\\
$M$ (MeV)&9490&9767&9780&9780&10052&10212&10215&10215&10292\\
$M^{\rm exp}$ (MeV)&9460&9860&9892&9913&10230&10235&10255&10269&10355\\
\hline
\hline
\end{tabular}

\end{center}

\end{table}

\begin{table}
\caption{The masses and weak decay constants (MeV) of heavy-light mesons.}
\label{hl}

\begin{center}

\begin{tabular}{lccccccccc}
\hline
\hline
{\small Meson }& $D$&$D^*$&$D_s$&$D_s^*$
&$B$&$B^*$&$B_s$&$B_s^*$\\
\hline
$M$ & 1766 & 1991 & 1910 & 2142 & 4965 & 5143 & 5092 & 5292 \\
$M^{\rm exp}$ & 1869 & 2010 & 1969 & 2110 & 5278 & 5324 & 5375 & 5422\\
$f_P$ & 149 & - & 177 & - & 123 & - & 150 & - \\
\hline
\hline
\end{tabular}

\end{center}

\end{table}

\newpage

\unitlength=1.00mm
\special{em:linewidth 0.4pt}
\linethickness{0.4pt}
\begin{picture}(41.00,145.00)
\put(35.00,140.00){\circle{10.00}}
\put(35.00,135.00){\circle*{4.00}}
\put(35.00,130.00){\circle*{4.00}}
\put(30.00,123.00){\line(3,4){5.33}}
\put(35.33,130.00){\line(4,-5){5.67}}
\end{picture}

\vspace{-12cm}

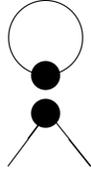
FIG.1. Divergent bubble diagram.

\unitlength=1.00mm
\special{em:linewidth 0.4pt}
\linethickness{0.4pt}
\begin{picture}(122.00,133.00)
\put(25.00,121.00){\circle{14.00}}
\put(61.00,120.00){\circle{14.00}}
\put(104.00,120.00){\circle{14.00}}
\put(18.00,121.00){\circle*{4.00}}
\put(32.00,121.00){\circle*{4.00}}
\put(56.00,125.00){\circle*{4.00}}
\put(64.00,126.00){\circle{4.00}}
\put(64.00,126.00){\circle*{4.00}}
\put(67.00,117.00){\circle*{4.00}}
\put(62.00,113.00){\circle*{4.00}}
\put(55.00,117.00){\circle*{4.00}}
\put(97.00,120.00){\circle*{4.00}}
\put(111.00,120.00){\circle*{0.00}}
\put(8.00,122.00){\line(1,0){10.00}}
\put(18.00,122.00){\line(0,-1){2.00}}
\put(18.00,120.00){\line(-1,0){10.00}}
\put(8.00,120.00){\line(0,0){0.00}}
\put(41.00,122.00){\line(-1,0){9.00}}
\put(32.00,122.00){\line(0,-1){2.00}}
\put(32.00,120.00){\line(1,0){9.00}}
\put(88.00,121.00){\line(1,0){9.00}}
\put(97.00,121.00){\line(0,-1){2.00}}
\put(97.00,119.00){\line(-1,0){9.00}}
\put(55.00,125.00){\line(-6,5){7.00}}
\put(56.00,126.00){\line(-6,5){7.00}}
\put(69.00,133.00){\line(-5,-6){5.67}}
\put(63.33,126.00){\line(5,-3){1.67}}
\put(65.00,125.00){\line(5,6){5.67}}
\put(74.00,113.00){\line(-3,2){6.00}}
\put(68.00,117.00){\line(-1,-1){1.00}}
\put(67.00,116.00){\line(3,-2){6.00}}
\put(63.00,104.00){\line(0,1){9.00}}
\put(63.00,113.00){\line(-1,0){2.00}}
\put(61.00,113.00){\line(0,-1){9.00}}
\put(50.00,111.00){\line(1,1){5.00}}
\put(55.00,116.00){\line(-1,1){1.00}}
\put(54.00,117.00){\line(-5,-4){5.00}}
\put(52.00,106.00){\makebox(0,0)[cc]{b}}
\put(93.00,106.00){\makebox(0,0)[cc]{c}}
\put(15.00,106.00){\makebox(0,0)[cc]{a}}
\put(115.00,118.00){\line(-2,1){4.00}}
\put(111.00,120.00){\line(2,1){4.00}}
\put(117.00,117.00){\line(2,-1){4.00}}
\put(117.00,123.00){\line(5,2){5.00}}
\end{picture}

\vspace{-7cm}

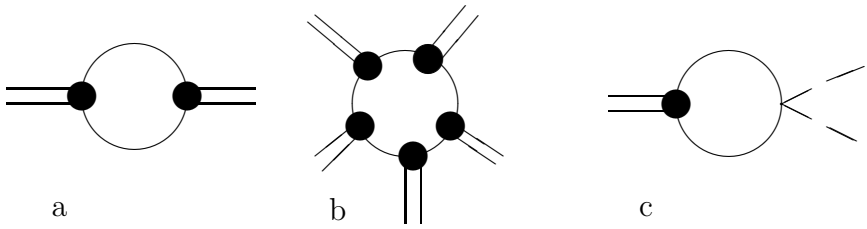
FIG.2. Diagrams describing different processes in effective nonlocal
meson theory in the lowest (one-loop) order.

\unitlength=1.00mm
\special{em:linewidth 0.4pt}
\linethickness{0.4pt}
\begin{picture}(106.00,146.00)
\put(10.00,145.00){\line(2,-1){10.00}}
\put(20.00,140.00){\line(-2,-1){10.00}}
\put(10.00,135.00){\line(0,0){0.00}}
\put(50.00,145.00){\line(-2,-1){10.00}}
\put(40.00,140.00){\line(2,-1){10.00}}
\put(89.00,140.00){\circle*{4.47}}
\put(95.00,140.00){\circle*{5.20}}
\put(89.00,140.00){\circle*{5.20}}
\put(79.00,145.00){\line(2,-1){11.00}}
\put(90.00,139.33){\line(-5,-2){11.00}}
\put(106.00,145.00){\line(-2,-1){12.00}}
\put(94.00,139.00){\line(3,-1){12.00}}
\put(53.00,140.00){\makebox(0,0)[lc]{~{\large =}~~~~$\sum_{aJ\ell nj}$}}
\put(90.00,133.00){\makebox(0,0)[rc]{$aJ\ell nj$}}
\put(94.00,146.00){\makebox(0,0)[lc]{$aJ\ell nj$}}
\put(38.00,138.00){\line(0,1){0.00}}
\put(20.00,140.00){\line(1,0){2.00}}
\put(22.00,140.00){\line(0,1){0.00}}
\put(24.00,140.00){\line(1,0){3.00}}
\put(29.00,140.00){\line(1,0){3.00}}
\put(34.00,140.00){\line(1,0){2.00}}
\put(38.00,140.00){\line(1,0){2.00}}
\put(10.00,119.00){\line(1,0){3.00}}
\put(15.00,119.00){\line(1,0){3.00}}
\put(20.00,119.00){\line(1,0){3.00}}
\put(25.00,119.00){\line(1,0){3.00}}
\put(10.00,109.00){\line(5,-3){10.00}}
\put(20.00,103.00){\line(-2,-1){10.00}}
\put(10.00,98.00){\line(0,1){0.00}}
\put(33.00,119.00){\makebox(0,0)[lc]{{\small Gluon propagator in 
the vacuum field}}}
\put(25.00,103.00){\makebox(0,0)[lc]{{\small Local quark current}}}
\put(76.00,103.00){\circle*{4.47}}
\put(65.00,108.00){\line(2,-1){11.00}}
\put(76.00,102.33){\line(-5,-2){11.00}}
\put(82.00,103.00){\makebox(0,0)[lc]{{\small Nonlocal quark currents}}}
\end{picture}

\vspace{-7cm}

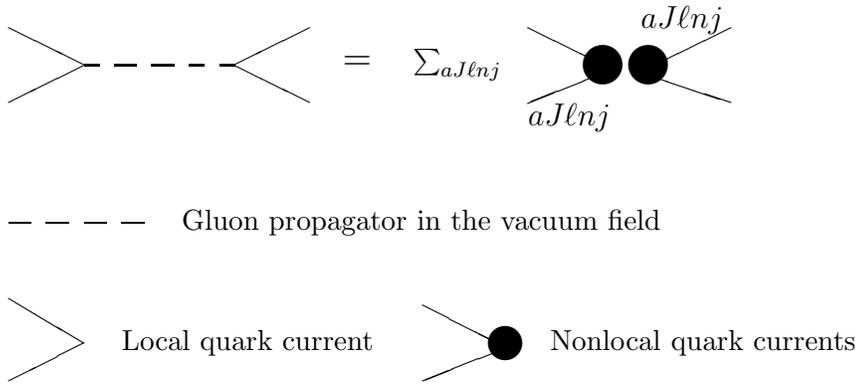
FIG.3. The decomposition of the four-quark interaction.

\end{document}